\documentclass[zpreprint,zbstpl]{zeus_paper}
\usepackage[english]{babel}
\usepackage{tikz}
\usetikzlibrary{positioning}
\usetikzlibrary{shapes.misc} % rounded rectangles, etc
\usetikzlibrary{shapes.arrows}
\usetikzlibrary{arrows}
\newcommand{\be}{\begin{equation}} %these definitions save typing
\newcommand{\ee}{\end{equation}}
\newcommand{\bea}{\begin{eqnarray}}
\newcommand{\eea}{\end{eqnarray}}

\newcommand{\ddif}[3]{\frac{d^{2}#1}{d#2 d#3}}
\usepackage[mathlines,displaymath]{lineno}
\newcommand{\Zsttdesc}[1]{%
The STT consisted of 48 sectors of two different sizes. Each sector
contained 192 (small sector) or 264 (large sector) straws of diameter
7.5 mm arranged into 3 layers. The sectors were trapezoidal in shape
and each subtended an azimuthal angle of $60^{\circ}$ -- 6 sectors
formed a so-called superlayer. A particle passing through the complete
detector traversed 8 superlayers, which were rotated around the beam
direction at angles of $30^{\circ}$ or $15^{\circ}$ to each other. The STT
covered the polar-angle region $5^{\circ}<\theta<23^{\circ}$.
}
\newcommand{\Zacknowledge}{%
We appreciate the contributions to the construction, maintenance 
and operation of the ZEUS detector of many people who are not listed 
as authors. The HERA 
machine group and the DESY computing staff are especially acknowledged for 
their success in providing excellent operation of the collider and the 
data-analysis environment. We thank the DESY directorate for their strong 
support and encouragement.}
\chardef\usc=95
\chardef\til=126
\catcode`\@=11 % @ signs are now treated as letters
\DeclareRobustCommand\xdotspace{\futurelet\@let@token\@xdotspace}
\def\@xdotspace{%
  \ifx\@let@token.\else
  \ifx\@let@token\bgroup.\else
  \ifx\@let@token\egroup.\else
  \ifx\@let@token\/.\else
  \ifx\@let@token\ .\else
  \ifx\@let@token~.\else
  \ifx\@let@token!.\else
  \ifx\@let@token,.\else
  \ifx\@let@token:.\else
  \ifx\@let@token;.\else
  \ifx\@let@token?.\else
  \ifx\@let@token/.\else
  \ifx\@let@token'.\else
  \ifx\@let@token).\else
  \ifx\@let@token-.\else
  \ifx\@let@token\@xobeysp.\else
  \ifx\@let@token\space.\else
  \ifx\@let@token\@sptoken.\else
   .\space
   \fi\fi\fi\fi\fi\fi\fi\fi\fi\fi\fi\fi\fi\fi\fi\fi\fi\fi}
\catcode`\@=12 % @ signs are no longer letters

\newcommand{\stru}[2]{%
   \relax\ifmmode\hbox{\vrule height#1 depth#2 width0pt}%
   \else\vrule height#1 depth#2 width0pt\fi}
\newcommand{\Ronum}[1]{\uppercase\expandafter{\romannumeral#1}}
\newcommand{\ronum}[1]{\expandafter{\romannumeral#1}}
\DeclareRobustCommand{\LaTeXZ}{%
  \LaTeX\kern-.05em4\kern-.1em
  {\raisebox{-0.2ex}{$\scriptstyle\text{ZEUS}$}}\xspace}
\DeclareMathAlphabet{\mathbf}{OT1}{cmr}{bx}{sl}
\newcommand{\eVdist}{\kern-0.06667em}

\newcommand{\gev}{{\,\text{Ge}\eVdist\text{V\/}}}

\newcommand{\pbi}{\,\text{pb}^{-1}}

\newcommand{\slashfrac}[2]{%
  \raisebox{0.5ex}{\ensuremath #1}\kern-0.12em/\kern-0.08em
  \raisebox{-.8ex}{\ensuremath #2}}
\newcommand{\sqr}[3]{%
    {\vcenter{\hrule height.#3ex\hbox{\vrule width.#2ex height#1ex
     \kern#1ex\vrule width.#3ex}\hrule height.#2ex}}}

\catcode`\@=11 % @ signs are now treated as letters
\newcommand{\parenbar}{\mathpalette\p@renb@r}
\def\p@renb@r#1#2{\vbox{%
  \ifx#1\scriptscriptstyle \dimen@.7em\dimen@ii.2em\else
  \ifx#1\scriptstyle \dimen@.8em\dimen@ii.25em\else
  \dimen@1em\dimen@ii.4em\fi\fi \offinterlineskip
  \ialign{\hfill##\hfill\cr
    \vbox{\hrule width\dimen@ii}\cr
    \noalign{\vskip-.3ex}%
    \hbox to\dimen@{$\mathchar300\hfil\mathchar301$}\cr
    \noalign{\vskip-.3ex}%
    $#1#2$\cr}}}
\catcode`\@=12 % @ signs are no longer letters

\newcommand{\IP}{{\rm I$\kern-0.01667em$P}\xspace}

\mathchardef\qsm=63
\mathchardef\pls=43
\mathchardef\mns=512
\mathchardef\plm=518
\mathchardef\eql=61
\mathchardef\smallleft=300
\mathchardef\smallright=301
\mathchardef\les=316
\mathchardef\gre=318
\mathchardef\leq=532
\mathchardef\grq=533
\catcode`\@=11 % @ signs are now treated as letters
\newcounter{pict@width}
\newcounter{pict@height}
\newlength{\pict@scale}
\newcommand{\psfigadd}[4]{%
\setcounter{pict@width}{1*\ratio{#2+\pict@scale/2}{\pict@scale}}
\setcounter{pict@height}{1*\ratio{#3+\pict@scale/2}{\pict@scale}}
\setlength{\unitlength}{\pict@scale}
\hbox to #2{\hspace{-\fill}\begin{picture}(\thepict@width,\thepict@height)
\put(0,0){\psfig{figure=#1,width=#2,height=#3,clip=}}
\SetScale{0.283466457}
\SetWidth{1.763889}
{#4}
\end{picture}}
}
\newcounter{pict@widthfst}
\newcounter{pict@widthscd}
\newcounter{pict@widthtot}
\newcommand{\psfigaddtwo}[7]{%
\setcounter{pict@widthfst}{1*\ratio{#2+\pict@scale/2}{\pict@scale}}
\setcounter{pict@widthscd}{1*\ratio{#2+#4+\pict@scale/2}{\pict@scale}}
\setcounter{pict@widthtot}{1*\ratio{#2+#4+#6+\pict@scale/2}{\pict@scale}}
\setcounter{pict@height}{1*\ratio{#3+\pict@scale/2}{\pict@scale}}
\setlength{\unitlength}{\pict@scale}
\hbox{\hspace{-\fill}\begin{picture}(\thepict@widthtot,\thepict@height)
\put(0,0){\psfig{figure=#1,width=#2,height=#3,clip=}}
\put(\thepict@widthscd,0){\psfig{figure=#5,width=#6,height=#3,clip=}}
\SetScale{0.283466457}
\SetWidth{1.763889}
{#7}
\end{picture}}
}
\newcommand{\psfigror}[4]{%
\setcounter{pict@width}{1*\ratio{#2+\pict@scale/2}{\pict@scale}}
\setcounter{pict@height}{1*\ratio{#3+\pict@scale/2}{\pict@scale}}
\setlength{\unitlength}{\pict@scale}
\hbox{\begin{picture}(\thepict@width,\thepict@height)
\put(0,\thepict@height){\psfig{figure=#1,width=#3,height=#2,clip=,angle=270}}
\SetScale{0.283466457}
\SetWidth{1.763889}
{#4}
\end{picture}}
}
\newcommand{\psfigrol}[4]{%
\setcounter{pict@width}{1*\ratio{#2+\pict@scale/2}{\pict@scale}}
\setcounter{pict@height}{1*\ratio{#3+\pict@scale/2}{\pict@scale}}
\setlength{\unitlength}{\pict@scale}
\hbox{\begin{picture}(\thepict@width,\thepict@height)
\put(0,0){\psfig{figure=#1,width=#3,height=#2,clip=,angle=90}}
\SetScale{0.283466457}
\SetWidth{1.763889}
{#4}
\end{picture}}
}
\catcode`\@=12 % @ signs are no longer letters
\newlength\listtextwidth

\catcode`\@=11 % @ signs are now treated as letters
\newlength{\@tabfninsert}
\newlength{\@tabfnwidth}
\newcommand{\tabfootnote}[2]{%
  \setlength{\@tabfninsert}{0.8em}
  \setlength{\@tabfnwidth}{\textwidth}
  \addtolength{\@tabfnwidth}{-\@tabfninsert}
  \addtolength{\@tabfnwidth}{-0.4em}
  \noindent\makebox[\@tabfninsert][r]{\footnotesize$^{#1}$\hfil}\hfill%
  \parbox[t]{\@tabfnwidth}{\footnotesize #2\hfill}}
\catcode`\@=12 % @ signs are no longer letters
\begin{document}
%------------------------------------------------------------------------------
%       Title sheet
%------------------------------------------------------------------------------
\prepnum{{DESY--16--039}}
%\title{
%Simultaenous determination of\\ Parton Distribution Functions \\
%and the $\bold{Z}$ couplings to $\bold{u}$ and $\bold{d}$ quarks\\
%and the mass of the $\bold{W}$ \\
%}                                                       
\title{
Combined QCD and electroweak analysis\\ of HERA data\\
}                                                       
                    
\author{ZEUS Collaboration}
\date{March 2016}

\abstract{

A simultaneous fit of parton distribution functions (PDFs) and electroweak 
parameters to HERA data on deep inelastic scattering is presented. 
The input data are the neutral current and charged current  
inclusive cross sections 
which were previously used in the QCD analysis leading to 
the HERAPDF2.0 PDFs. In addition, 
the polarisation of the electron beam was taken into account 
for the ZEUS data recorded between 2004 and 2007. Results 
on the vector and axial-vector couplings of the $Z$ boson to 
$u$- and $d$-type quarks, on the value of the electroweak mixing angle 
and the mass of the $W$ boson are presented. The values obtained 
for the electroweak parameters are in agreement with 
Standard Model predictions. 
%The resulting sets of PDFs, ZEUS-EW, 
%are in agreement with HERAPDF2.0 and give a good description 
%of ZEUS data with polarisation taken into account. 

}

\makezeustitle

\pagenumbering{roman}
                                    % this "%"s are for cosmetics only

                                                   %
                                                   %
                                                   %
\begin{center}
{                      \Large  The ZEUS Collaboration              }
\end{center}

{\small\raggedright

%  members:

H.~Abramowicz$^{25, u}$, 
I.~Abt$^{20}$, 
L.~Adamczyk$^{8}$, 
M.~Adamus$^{31}$, 
S.~Antonelli$^{2}$, 
V.~Aushev$^{17}$, 
O.~Behnke$^{10}$, 
U.~Behrens$^{10}$, 
A.~Bertolin$^{22}$, 
S.~Bhadra$^{33}$
% ...
}
\newpage

%       institutes:

{\setlength{\parskip}{0.4em}
\makebox[3ex]{$^{1}$}
\begin{minipage}[t]{14cm}
{\it INFN Bologna, Bologna, Italy}~$^{A}$

\end{minipage}

\makebox[3ex]{$^{2}$}
\begin{minipage}[t]{14cm}
{\it University and INFN Bologna, Bologna, Italy}~$^{A}$

\end{minipage}

\makebox[3ex]{$^{3}$}
\begin{minipage}[t]{14cm}
{\it Physikalisches Institut der Universit\"at Bonn,
Bonn, Germany}~$^{B}$

\end{minipage}

\makebox[3ex]{$^{4}$}
\begin{minipage}[t]{14cm}
{\it Panjab University, Department of Physics, Chandigarh, India}

\end{minipage}

\makebox[3ex]{$^{5}$}
\begin{minipage}[t]{14cm}
{\it Calabria University,
Physics Department and INFN, Cosenza, Italy}~$^{A}$

\end{minipage}

\makebox[3ex]{$^{6}$}
\begin{minipage}[t]{14cm}
{\it National Centre for Particle Physics, Universiti Malaya, 50603 Kuala Lumpur, Malaysia}~$^{C}$

\end{minipage}

\makebox[3ex]{$^{7}$}
\begin{minipage}[t]{14cm}
{\it The Henryk Niewodniczanski Institute of Nuclear Physics, Polish Academy of \\
Sciences, Krakow, Poland}~$^{D}$

\end{minipage}

\makebox[3ex]{$^{8}$}
\begin{minipage}[t]{14cm}
{\it AGH-University of Science and Technology, Faculty of Physics and Applied Computer
Science, Krakow, Poland}~$^{D}$

\end{minipage}

\makebox[3ex]{$^{9}$}
\begin{minipage}[t]{14cm}
{\it Department of Physics, Jagellonian University, Krakow, Poland}

\end{minipage}

\makebox[3ex]{$^{10}$}
\begin{minipage}[t]{14cm}
{\it Deutsches Elektronen-Synchrotron DESY, Hamburg, Germany}

\end{minipage}

\makebox[3ex]{$^{11}$}
\begin{minipage}[t]{14cm}
{\it Deutsches Elektronen-Synchrotron DESY, Zeuthen, Germany}

\end{minipage}

\makebox[3ex]{$^{12}$}
\begin{minipage}[t]{14cm}
{\it School of Physics and Astronomy, University of Glasgow,
Glasgow, United Kingdom}~$^{E}$

\end{minipage}

\makebox[3ex]{$^{13}$}
\begin{minipage}[t]{14cm}
{\it Hamburg University, Institute of Experimental Physics, Hamburg,
Germany}~$^{F}$

\end{minipage}

\makebox[3ex]{$^{14}$}
\begin{minipage}[t]{14cm}
{\it Institute of Particle and Nuclear Studies, KEK,
Tsukuba, Japan}~$^{G}$

\end{minipage}

\makebox[3ex]{$^{15}$}
\begin{minipage}[t]{14cm}
{\it Institute of Physics and Technology of Ministry of Education and
Science of Kazakhstan, Almaty, Kazakhstan}

\end{minipage}

\makebox[3ex]{$^{16}$}
\begin{minipage}[t]{14cm}
{\it Institute for Nuclear Research, National Academy of Sciences, Kyiv, Ukraine}

\end{minipage}

\makebox[3ex]{$^{17}$}
\begin{minipage}[t]{14cm}
{\it Department of Nuclear Physics, National Taras Shevchenko University of Kyiv, Kyiv, Ukraine}

\end{minipage}

\makebox[3ex]{$^{18}$}
\begin{minipage}[t]{14cm}
{\it Meiji Gakuin University, Faculty of General Education,
Yokohama, Japan}~$^{G}$

\end{minipage}

\makebox[3ex]{$^{19}$}
\begin{minipage}[t]{14cm}
{\it Lomonosov Moscow State University, Skobeltsyn Institute of Nuclear Physics,
Moscow, Russia}~$^{H}$

\end{minipage}

\makebox[3ex]{$^{20}$}
\begin{minipage}[t]{14cm}
{\it Max-Planck-Institut f\"ur Physik, M\"unchen, Germany}

\end{minipage}

\makebox[3ex]{$^{21}$}
\begin{minipage}[t]{14cm}
{\it Department of Physics, University of Oxford,
Oxford, United Kingdom}~$^{E}$

\end{minipage}

\makebox[3ex]{$^{22}$}
\begin{minipage}[t]{14cm}
{\it INFN Padova, Padova, Italy}~$^{A}$

\end{minipage}

\makebox[3ex]{$^{23}$}
\begin{minipage}[t]{14cm}
{\it Dipartimento di Fisica e Astronomia dell' Universit\`a and INFN,
Padova, Italy}~$^{A}$

\end{minipage}

\makebox[3ex]{$^{24}$}
\begin{minipage}[t]{14cm}
{\it Polytechnic University, Tokyo, Japan}~$^{G}$

\end{minipage}

\makebox[3ex]{$^{25}$}
\begin{minipage}[t]{14cm}
{\it Raymond and Beverly Sackler Faculty of Exact Sciences, School of Physics, \\
Tel Aviv University, Tel Aviv, Israel}~$^{I}$

\end{minipage}

\makebox[3ex]{$^{26}$}
\begin{minipage}[t]{14cm}
{\it Department of Physics, Tokyo Institute of Technology,
Tokyo, Japan}~$^{G}$

\end{minipage}

\makebox[3ex]{$^{27}$}
\begin{minipage}[t]{14cm}
{\it Universit\`a di Torino and INFN, Torino, Italy}~$^{A}$

\end{minipage}

\makebox[3ex]{$^{28}$}
\begin{minipage}[t]{14cm}
{\it Universit\`a del Piemonte Orientale, Novara, and INFN, Torino,
Italy}~$^{A}$

\end{minipage}

\makebox[3ex]{$^{29}$}
\begin{minipage}[t]{14cm}
{\it Physics and Astronomy Department, University College London,
London, United Kingdom}~$^{E}$

\end{minipage}

\makebox[3ex]{$^{30}$}
\begin{minipage}[t]{14cm}
{\it Faculty of Physics, University of Warsaw, Warsaw, Poland}

\end{minipage}

\makebox[3ex]{$^{31}$}
\begin{minipage}[t]{14cm}
{\it National Centre for Nuclear Research, Warsaw, Poland}

\end{minipage}

\makebox[3ex]{$^{32}$}
\begin{minipage}[t]{14cm}
{\it Department of Particle Physics and Astrophysics, Weizmann
Institute, Rehovot, Israel}

\end{minipage}

\makebox[3ex]{$^{33}$}
\begin{minipage}[t]{14cm}
{\it Department of Physics, York University, Ontario, Canada M3J 1P3}~$^{J}$

\end{minipage}

}

\vspace{3em}

%  references concerning institutes;

{\setlength{\parskip}{0.4em}\raggedright
\makebox[3ex]{$^{ A}$}
\begin{minipage}[t]{14cm}
 supported by the Italian National Institute for Nuclear Physics (INFN) \
\end{minipage}

\makebox[3ex]{$^{ B}$}
\begin{minipage}[t]{14cm}
 supported by the German Federal Ministry for Education and Research (BMBF), under
 contract No. 05 H09PDF\
\end{minipage}

\makebox[3ex]{$^{ C}$}
\begin{minipage}[t]{14cm}
 supported by HIR grant UM.C/625/1/HIR/149 and UMRG grants RU006-2013, RP012A-13AFR and RP012B-13AFR from
 Universiti Malaya, and ERGS grant ER004-2012A from the Ministry of Education, Malaysia\
\end{minipage}

\makebox[3ex]{$^{ D}$}
\begin{minipage}[t]{14cm}
 supported by the National Science Centre under contract No. DEC-2012/06/M/ST2/00428\
\end{minipage}

\makebox[3ex]{$^{ E}$}
\begin{minipage}[t]{14cm}
 supported by the Science and Technology Facilities Council, UK\
\end{minipage}

\makebox[3ex]{$^{ F}$}
\begin{minipage}[t]{14cm}
 supported by the German Federal Ministry for Education and Research (BMBF), under
 contract No. 05h09GUF, and the SFB 676 of the Deutsche Forschungsgemeinschaft (DFG) \
\end{minipage}

\makebox[3ex]{$^{ G}$}
\begin{minipage}[t]{14cm}
 supported by the Japanese Ministry of Education, Culture, Sports, Science and Technology
 (MEXT) and its grants for Scientific Research\
\end{minipage}

\makebox[3ex]{$^{ H}$}
\begin{minipage}[t]{14cm}
 supported by RF Presidential grant N 3042.2014.2 for the Leading Scientific Schools\
\end{minipage}

\makebox[3ex]{$^{ I}$}
\begin{minipage}[t]{14cm}
 supported by the Israel Science Foundation\
\end{minipage}

\makebox[3ex]{$^{ J}$}
\begin{minipage}[t]{14cm}
 supported by the Natural Sciences and Engineering Research Council of Canada (NSERC) \
\end{minipage}

}

\pagebreak[4]
{\setlength{\parskip}{0.4em}

%  references concerning members;

\makebox[3ex]{$^{ a}$}
\begin{minipage}[t]{14cm}
now at INFN Roma, Italy\
\end{minipage}

\makebox[3ex]{$^{ b}$}
\begin{minipage}[t]{14cm}
now at Sant Longowal Institute of Engineering and Technology, Longowal, Punjab, India\
\end{minipage}

\makebox[3ex]{$^{ c}$}
\begin{minipage}[t]{14cm}
now at Sri Guru Granth Sahib World University, Fatehgarh Sahib, India\
\end{minipage}

\makebox[3ex]{$^{ d}$}
\begin{minipage}[t]{14cm}
also at Agensi Nuklear Malaysia, 43000 Kajang, Bangi, Malaysia\
\end{minipage}

\makebox[3ex]{$^{ e}$}
\begin{minipage}[t]{14cm}
partially supported by the Polish National Science Centre projects DEC-2011/01/B/ST2/03643 and DEC-2011/03/B/ST2/00220\
\end{minipage}

\makebox[3ex]{$^{ f}$}
\begin{minipage}[t]{14cm}
now at Rockefeller University, New York, NY 10065, USA\
\end{minipage}

\makebox[3ex]{$^{ g}$}
\begin{minipage}[t]{14cm}
now at European X-ray Free-Electron Laser facility GmbH, Hamburg, Germany\
\end{minipage}

\makebox[3ex]{$^{ h}$}
\begin{minipage}[t]{14cm}
now at University of Liverpool, United Kingdom\
\end{minipage}

\makebox[3ex]{$^{ i}$}
\begin{minipage}[t]{14cm}
now at Tel Aviv University, Isreal\
\end{minipage}

\makebox[3ex]{$^{ j}$}
\begin{minipage}[t]{14cm}
now at Physikalisches Institut, Universit\"{a}t Heidelberg, Germany\
\end{minipage}

\makebox[3ex]{$^{ k}$}
\begin{minipage}[t]{14cm}
supported by the Alexander von Humboldt Foundation\
\end{minipage}

\makebox[3ex]{$^{ l}$}
\begin{minipage}[t]{14cm}
now at CERN, Geneva, Switzerland\
\end{minipage}

\makebox[3ex]{$^{ m}$}
\begin{minipage}[t]{14cm}
Alexander von Humboldt Professor; also at DESY and University of Oxford\
\end{minipage}

\makebox[3ex]{$^{ n}$}
\begin{minipage}[t]{14cm}
also at DESY\
\end{minipage}

\makebox[3ex]{$^{ o}$}
\begin{minipage}[t]{14cm}
also at University of Tokyo, Japan\
\end{minipage}

\makebox[3ex]{$^{ p}$}
\begin{minipage}[t]{14cm}
now at Kobe University, Japan\
\end{minipage}

\makebox[3ex]{$^{ q}$}
\begin{minipage}[t]{14cm}
member of National Technical University of Ukraine, Kyiv Polytechnic Institute, Kyiv, Ukraine\
\end{minipage}

\makebox[3ex]{$^{ r}$}
\begin{minipage}[t]{14cm}
now at DESY CMS group\
\end{minipage}

\makebox[3ex]{$^{ s}$}
\begin{minipage}[t]{14cm}
now at DESY ATLAS group\
\end{minipage}

\makebox[3ex]{$^{ t}$}
\begin{minipage}[t]{14cm}
now at LNF, Frascati, Italy\
\end{minipage}

\makebox[3ex]{$^{ u}$}
\begin{minipage}[t]{14cm}
also at Max Planck Institute for Physics, Munich, Germany, External Scientific Member\
\end{minipage}

\makebox[3ex]{$^{ v}$}
\begin{minipage}[t]{14cm}
also supported by DESY and the Alexander von Humboldt Foundation\
\end{minipage}

\makebox[3ex]{$^{ w}$}
\begin{minipage}[t]{14cm}
also at \L\'{o}d\'{z} University, Poland\
\end{minipage}

\makebox[3ex]{$^{ x}$}
\begin{minipage}[t]{14cm}
member of \L\'{o}d\'{z} University, Poland\
\end{minipage}

\makebox[3ex]{$^{ y}$}
\begin{minipage}[t]{14cm}
now at Polish Air Force Academy in Deblin\
\end{minipage}

}

%%%% \begin{center}
%%%% {                      \Large  The ZEUS Collaboration              }
%%%% \end{center}

%{\small

%  members:

\newpage

%------------------------------------------------------------------------------
%       Text
%------------------------------------------------------------------------------
\pagenumbering{arabic} 
\pagestyle{plain}

% ----------------------------------------------------------------------------
%       Introduction
% ----------------------------------------------------------------------------
\section{Introduction}
\label{sec-int}
Data on deep inelastic scattering (DIS) of leptons from nucleons 
have been used for many years in many ways to test the 
Standard Model (SM) of the electroweak and strong 
interactions~\cite{MANDY} and have been fundamental 
in unravelling the structure of nucleons.
The %first and only 
electron--proton, $ep$, collider HERA extended the reach in the 
four-momentum-transfer squared, $Q^2$, and in Bjorken $x$ by 
several orders of magnitude with respect to previous fixed-target 
experiments~\cite{rmp:71:1275}. 
At HERA,
the values of $Q^2$ extend up to $50\,000\,\gev^2$, 
where the $Z$-exchange contribution is comparable 
to that of the photon exchange.
This, together with the longitudinal polarisation of the 
electrons\,\footnote{In this paper, the word ``electron'' 
refers to both electrons and positrons, unless otherwise stated.} 
in the beam, have made 
a significant test of the couplings of the $Z$ to the
quarks possible. 
The on-shell value of the
electroweak mixing angle, $\sin^2\theta_W$, 
and of the mass of the $W$ boson, $M_W$,
were also determined via a combined QCD and electroweak analysis.

The HERA collider was operated in two phases, 
HERA\,I: 1992--2000 and HERA\,II: 2003--2007.
During the HERA\,II phase, the electron beams were longitudinally
polarised to a level between 25\,\% and 35\,\%. 
A combination of all ZEUS and H1 inclusive data for zero polarisation was 
published %~\cite{HERAPDF20}
and subject to a detailed QCD analysis~\cite{HERAPDF20}, 
yielding the parton distribution function (PDF) set HERAPDF2.0 and 
its variants.
For the analysis presented here,
the ZEUS HERA\,II data taken at the
centre-of-mass energy of 318\,GeV
were used separated into sets with positive and 
negative polarisation
as published by the ZEUS
collaboration~\cite{ZEUS2NCe, ZEUS2NCp, ZEUS2CCe, ZEUS2CCp}. 
All other data sets were used as
originally published by
H1~\cite{H1allhQ2,H1FL1,H1FL2,Adloff:1999ah,Adloff:2000qj,Adloff:2003uh,Collaboration:2009kv,Collaboration:2009bp}
and ZEUS~\cite{ZEUSFL,Breitweg:1998dz,zeuscc97,Chekanov:2001qu,Chekanov:2002zs,Chekanov:2002ej,Chekanov:2003vw,Chekanov:2003yv} for unpolarised beams.
%\cite{HERAPDF20,HERAIcombi} and references therein.

\section{Standard Model formalism}\label{sec-pred}

Inclusive deep inelastic $ep$ scattering can be described in
terms of the kinematic variables $Q^2$, $x_{\rm{Bj}}$ and $y$.  
The negative four-momentum-transfer squared, $Q^2$,
is defined as $Q^2 = -q^2 = -(k-k')^2$, 
where $k$ and $k'$ are the four-momenta of the incoming and 
the scattered electron, respectively.
The Bjorken scaling variable, $x_{\rm{Bj}}$, is defined 
as $x_{\rm{Bj}}=Q^2/2P \cdot q$, where $P$ is
the four-momentum of the incoming proton. 
In the quark-parton model (QPM)  
the kinematic variable  $x_{\rm{Bj}}$ is equal to the fractional
momentum of the struck quark, $x$.
The fraction of the electron energy transferred to the proton in the rest frame
of the proton is given by $y = P \cdot q / P \cdot k$. 
At HERA energies, the masses of the incoming electrons (protons)
with energies $E_e$ ($E_p$) can be neglected and
the variables $Q^2$, $x_{\rm{Bj}}$ and $y$
%$x_{\rm{Bj}}$, $y$ and $Q^2$ 
are related as $Q^2=sx_{\rm{Bj}}y$, 
where  $s=4E_eE_p$ is the square of the electron--proton centre-of-mass energy. 
%At HERA, $s=4E_e E_p$, where $E_{e}$ and $E_{p}$ 
%are the initial electron and proton energies, respectively. 

The components of the Standard Model necessary to describe the data are 
the electroweak (EW) theory  and
perturbative Quantum Chromo Dynamics (pQCD). 
At leading order, the EW theory supplies the cross sections
for electron scattering 
from partons with electric charge, i.e.\ the quarks. 
The EW theory is subject to pQCD corrections, which 
already at next-to-leading order make the electron scattering sensitive
to the gluons in the proton.
The dynamics of the partons, quarks and gluons,
are described via their PDFs.
The PDFs provide
the probability of finding a given parton with a momentum fraction $x$
for an interaction at a given 
factorisation scale, $\mu_{\rm f}$, which is usually chosen to be $Q^2$. 
In pQCD, the PDFs evolve with $Q^2$
depending on the order of the strong coupling constant, $\alpha_s$,
at which the perturbative series is truncated.
The analysis presented in this paper was performed at 
next-to-leading order (NLO) in pQCD.

The $ep$ cross sections measured at HERA 
were  published after they were
corrected for leading order (LO) quantum-electrodynamic (QED) radiative 
effects.
%~\footnote{The historical
%motivation for this was the removal of QED radiative effects, 
%such that a pure QCD analysis like the
%one for HERAPDF2.0 could be performed without 
%the need to explicitly consider QED corrections in the theory.
%}.
These are dominated by initial- and final-state photon emission 
by the electron.

The neutral current (NC) cross section 
at all orders of pQCD for  $e^ \pm p$ 
scattering can be written as~\cite{MANDY}
\begin{linenomath*}\begin{equation}
  \ddif{\sigma_{\rm NC}(e^{\pm}p)}{x_{\rm Bj}}{Q^{2}} = 
  \frac{2 \pi \alpha^{2} }{x_{\rm Bj}Q^{4}}
  [Y_{+} \, \tilde{F_{2}}(x_{\rm Bj},Q^{2})  
  \mp Y_{-} \, x\tilde{F_{3}}(x_{\rm Bj},Q^{2})  
  - y^{2}\tilde {F_{L}}(x_{\rm Bj},Q^{2})],
\label{eqn:unpol_xsec}
\end{equation}\end{linenomath*}
where $\alpha$ is the fine-structure constant,
$Y_{\pm} = 1 \pm (1 - y)^{2}$ and
$\tilde{F_{2}}(x_{\rm Bj},Q^{2})$, $x\tilde{F_{3}}(x_{\rm Bj},Q^{2})$ and
$\tilde{F_{L}}(x_{\rm Bj},Q^{2})$
are generalised structure functions. 
%Brian wants this 
The sign in front of the $x\tilde{F_3}$ term is taken as positive for 
electrons and negative for positrons.

The $\tilde{F_{2}}$ term in Eq.~\ref{eqn:unpol_xsec} is dominant at 
low $Q^{2}$, where only the photon exchange is important.
The longitudinal structure function $\tilde{F_{L}}$ is only
significant at very low  $Q^{2}$ and irrelevant for this analysis.
The $x\tilde{F_{3}}$ term starts to contribute significantly 
to the cross section at  
$Q^{2}$ values approaching the mass of the $Z$-boson squared, $M_Z^2$.
The latter originates from $\gamma / Z$ interference and $Z$ exchange
and results in a 
decrease (increase) of the $e^{+} p$ ($e^{-} p$) 
cross sections, respectively. 

The data were published as reduced cross sections
which were defined
for $e^-p$ and $e^+p$ NC scattering as
\begin{linenomath*}\begin{equation}
  \sigma^{e^{\pm} p}_{r,{\rm NC}} 
  =
  \frac {x_{\rm Bj}Q^{4}} {2 \pi \alpha^{2}_0}
  \frac {1} {Y_{+}}
  \ddif{\sigma(e^{\pm}p)}{x_{\rm Bj}}{Q^{2}}
  =
  \tilde{F_{2}}(x_{\rm Bj},Q^{2}) \mp \frac {Y_{-}} {Y_{+}} x \tilde{F_{3}}
  (x_{\rm Bj},Q^{2})- \frac {y^2} {Y_{+}} F_{L}(x_{\rm Bj},Q^{2}).
\label{eqn:red}
\end{equation}\end{linenomath*}
In this definition, 
the fine-structure constant is fixed to $\alpha_0$, i.e.\,at 
scale zero.
The QED corrections applied to the data use a 
running $\alpha$ to correct the data accordingly.

The generalised structure functions depend on the longitudinal polarisation 
of the electron beam, which is defined as
\begin{linenomath*}
\begin{equation}
P_{e}=\frac{N_{R}-N_{L}}{N_{R}+N_{L}}, 
%\nonumber 
\label{eqn:pol}
\end{equation}
\end{linenomath*}
where $N_{R}$ and $N_{L}$ are the numbers of right- 
and left-handed electrons in
the beam.

In all orders, the functions $\tilde{F_2}^\pm$ and $x\tilde{F_3}^\pm$  
can be split into structure-function terms depending on $\gamma$
exchange ($F_2^{\gamma}$), $Z$ exchange ($F_2^Z$, $xF_3^Z$) 
and $\gamma/Z$ interference ($F_2^{\gamma Z}$, $xF_3^{\gamma Z}$) as
\begin{linenomath*}\begin{equation}
    \tilde{F_2}^\pm = F_2^{\gamma} - (v_e \pm P_e a_e) \chi_{Z} F_2^{\gamma Z} +
  (v_e^2 + a_e^2 \pm 2 P_e v_e a_e) {\chi_{Z}^{2}} F_2^{Z}  ,
\label{eqn:gen_f2}
\end{equation}\end{linenomath*}
\begin{linenomath*}\begin{equation}
   x\tilde{F_3}^\pm =  - (a_e \pm P_e v_e) \chi_{Z} xF_3^{\gamma Z} + (2 v_e a_e 
\pm P_e(v_e^2 + a_e^2)) {\chi_{Z}^{2}} xF_3^{Z}~,
\label{eqn:gen_xf3}
\end{equation}\end{linenomath*}
where $\chi_{Z}$ is the relative strength of $Z$ exchange with respect to photon exchange.
These structure functions depend on the vector and axial-vector couplings
of the $Z$ boson to the electron.
The SM predictions for these couplings are 
$v_{e} = -1/2 + 2\sin^2\theta_W$ and $a_{e} = -1/2$. 
The on-shell definition of 
$\sin^2\theta_W = 1 - M_W^2/M_Z^2$ was chosen for the analysis.
In the on-shell scheme, this definition is valid to all orders
and $M_W$ becomes
\begin{linenomath*}\begin{equation}
M_W = \frac{A_0}{\sin^2\theta_W \sqrt{(1 - \Delta R)}} ~,
\label{eq:rewrite1}
\end{equation}\end{linenomath*}
where $A_0=\sqrt{\pi \alpha_0 / \sqrt{2} {G_{F}}} = 37.28039\,$GeV is a 
constant~\cite{PDG14}, $G_{F}$ is the Fermi coupling constant and $\Delta R$
accounts for radiative corrections, the running of
$\alpha$ and bosonic loop corrections dominated by the influence
of the mass of the top quark~\cite{PDG14}.

The relative strength of $Z$ exchange with respect to $\gamma$ 
exchange  depends on the on-shell $\sin^2\theta_W$ and
$\Delta R$  as
\begin{linenomath*}\begin{equation}
\chi_{Z}=\frac{1}{\sin^2{2\theta_W}} \frac{Q^{2}}{M_{Z}^{2}+Q^{2}} \frac{1}{1 - \Delta R} ~,
  \label{eqn:chiz}
\end{equation}\end{linenomath*}
where $M_Z$ is the pole mass of the $Z$ boson.
The value of $\chi_Z$ is 0.03 at $Q^2 = 185\,$GeV$^2$,
%Brians wishes
the lowest value of $Q^2$ for which ZEUS published 
inclusive NC cross sections with polarised beams
and increases to
1.1 at $Q^2 = 50\,000\,$GeV$^2$. 
Since from Eqs.~\ref{eqn:gen_f2} and~\ref{eqn:gen_xf3} polarisation only 
enters the structure functions via terms proportional 
to $\chi_Z$ or $\chi_Z^2$,
it is evident that beam polarisation 
predominantly affects the cross sections at high $Q^2$. 

Although this analysis was performed at NLO in QCD, 
the dominant contributions of the data sets can 
be identified by considering the structure functions in the framework 
of the QPM.
In this framework, the 
structure functions can be written in terms of sums and differences 
of the quark and anti-quark PDFs as 

\begin{linenomath*}\begin{equation}
[F_2^{\gamma},F_2^{\gamma Z},F_2^{Z}] = 
\sum _q [e_{q}^{2}, 2e_{q}v_{q},v_{q}^{2}+a_{q}^{2}] 
 x (q + \bar{q}),
\label{eqn:struc1}
\end{equation}\end{linenomath*}
\begin{linenomath*}\begin{equation}
[xF_3^{\gamma Z},xF_3^{Z}] = 
\sum _q [e_{q}a_{q},v_{q}a_{q}] 
 2x (q - \bar{q}),
\label{eqn:struc2}
\end{equation}\end{linenomath*}
where $v_{q}$ and $a_{q}$ are the respective vector and axial-vector couplings 
of the quark $q$ to the $Z$ boson, 
and $e_{q}$ is the electric charge of the quark. 
The PDFs of the quarks and anti-quarks are denoted $q$ and $\bar{q}$,
respectively. 
%In the QPM, only $u$ and $d$ quarks are present in the proton.

At any order in pQCD,
all quarks kinematically accessible
at HERA, i.e.\ all quarks except the top quark, have to be considered
in Eqs.~\ref{eqn:struc1} and~\ref{eqn:struc2},
but the sums are dominated
by $u$- and $d$-quark contributions. 
It is assumed throughout the analysis that all $u$-type
quarks have the same couplings, as do all
$d$-type quarks.
The SM predictions for the couplings are 
$v_{u} = 1/2 - 4/3 \sin^2\theta_W$, $a_{u} = 1/2$ and 
$v_{d} = -1/2 + 2/3 \sin^2\theta_W$, $a_{d} = -1/2$.

For most of the HERA phase space, $\chi^2_Z \ll \chi_Z$ and thus the
influence of pure $Z$ exchange is small. In addition, 
$v_e \approx 0.04$ is small. Thus
in Eqs.~\ref{eqn:gen_f2} and~\ref{eqn:gen_xf3} 
the axial-vector couplings
are determined predominantly through the term 
$-a_e \chi_Z xF_3^{\gamma Z}$ and the vector couplings through the term 
$-P_e a_e  \chi_Z F_2^{\gamma Z}$.
Thus the data obtained with polarised electron beams are crucial for 
a precise determination of the vector couplings.
Nevertheless, it is the combination of all data that provides
the final precision.

The charged current (CC) cross sections 
provide direct information on $M_W$.
Taking polarisation into account, they can be written as
\begin{linenomath*}\begin{equation}
  \ddif{\sigma_{\rm CC}(e^+p)}{x_{\rm Bj}}{Q^{2}} = (1 + P_e)
  \frac{G_F^{2}M_W^4 }{2\pi x_{\rm Bj}(Q^2 + M_W^2)^{2}}
  \,x\, [ (\bar{u} + \bar{c}) + (1 -y)^2 (d + s + b)]\,,
\label{eqn:xsec_CC1}
\end{equation}\end{linenomath*}

\begin{linenomath*}\begin{equation}
  \ddif{\sigma_{\rm CC}(e^-p)}{x_{\rm Bj}}{Q^{2}} = (1 - P_e)
  \frac{G_F^{2}M_W^4 }{2\pi x_{\rm Bj}(Q^2 + M_W^2)^{2}}
  \,x\, [ (u + c) + (1 -y)^2 (\bar{d} + \bar{s} + \bar{b})]\,.
\label{eqn:xsec_CC2}
\end{equation}\end{linenomath*}
It follows from Eq.~\ref{eq:rewrite1}
that the coupling $G_F$ can be rewritten in terms
of $\sin^2\theta_W$ and $M_W$ as
\begin{linenomath*}\begin{equation}
  G_F = \frac{\pi \alpha_0}{\sqrt{2}\, \sin^2\theta_W \,M_W^2}\, 
        \frac{1}{1 - \Delta R}~.
\label{eqn:GF}
\end{equation}\end{linenomath*}
Substituting $G_F$  into Eqs.~\ref{eqn:xsec_CC1}
and~\ref{eqn:xsec_CC2} parameterises 
the dependence of the CC cross sections on
$\sin^2\theta_W$.

%where $\frac{1}{1 - \Delta R}$ is the same electroweak correction factor 
%connected to the on-shell 
%scheme as used in Eq.~\ref{eqn:chiz}.

% ----------------------------------------------------------------------------
%       Experimental set-up
% ----------------------------------------------------------------------------
\section{Experimental setup}
\label{sec-setup}
% changed after group draft

The analysis is based on  inclusive cross sections for $ep$ scattering
published by the H1~\cite{H1allhQ2,H1FL1,H1FL2,Adloff:1999ah,Adloff:2000qj,Adloff:2003uh,Collaboration:2009kv,Collaboration:2009bp}
and ZEUS~\cite{ZEUS2NCe, ZEUS2NCp, ZEUS2CCe, ZEUS2CCp,ZEUSFL,Breitweg:1998dz,zeuscc97,Chekanov:2001qu,Chekanov:2002zs,Chekanov:2002ej,Chekanov:2003vw,Chekanov:2003yv}
collaborations for both the HERA\,II and
HERA\,I periods. 
A description 
%and complete list 
of all data sets, 
%used in the analysis presented here, 
including their respective integrated
luminosities was published previously~\cite{HERAPDF20}. 
All data sets were taken as input individually;
data sets were not combined, in contrast to the HERAPDF2.0 analysis.

Polarised beams were available for the HERA\,II period
from 2003 to 2007 when the electron beam energy was
$E_e = 27.5\,\gev$ and the proton beam was $E_p = 920\,\gev$, 
corresponding to a centre-of-mass energy of 318\,GeV.
The information on beam polarisation was used
in this analysis for the corresponding ZEUS HERA\,II data sets;
the H1 HERA\,II data sets were used as published for
zero polarisation.
%but not for the corresponding H1 data sets~\cite.
The kinematic range of these ZEUS HERA\,II data, see Table~\ref{tab:pol},
is $185 < Q^2 < 51\,200$\,GeV$^2$, $0.0063 < x_{\rm Bj} < 0.75$ for NC
and $200 < Q^2 < 60\,000$\,GeV$^2$, $0.0078  < x_{\rm Bj} < 1.0$ for CC
interactions.
 
The electron beam in HERA became naturally transversely polarised
through the Sokolov-Ternov effect~\cite{sovpdo:8:1203}.
The characteristic build-up time in HERA
was approximately 40 minutes.
Spin rotators on either side of the ZEUS detector
changed the transverse polarisation of the beam
into longitudinal polarisation in front of the interaction point 
and subsequently back to transverse polarisation.
The electron-beam polarisation was measured
using two independent polarimeters,
the transverse polarimeter (TPOL)~\cite {Baier:1969hw,nim:a329:79} and
the longitudinal polarimeter (LPOL)~\cite {nim:a479:334}.
Both devices exploited the spin-dependent cross section
for Compton scattering of circularly polarised photons from electrons.
%to measure the beam polarisation. / Brian
The luminosity and polarisation measurements were made over time scales
that were much shorter than the polarisation build-up time.

The total integrated luminosity for the ZEUS HERA\,II samples is 
about $300\,\pbi$. The data were almost evenly divided
between positive and negative beam polarisation.
The ZEUS %%%%%collaboration published the /Brian
cross sections for polarised electron beams were published
previously~\cite{ZEUS2NCe, ZEUS2NCp, ZEUS2CCe, ZEUS2CCp}.
For this analysis, the polarisation values were 
corrected using the final information on the polarimeters~\cite{Polanew}.
The relevant data sets and their polarisation values are
listed in Table~\ref{tab:pol}. The polarisation values do 
not differ by more than
0.3\,\% from  the previously published values for any
data set. 
%{\red do we want to publish LPOL and TPOL fractions? -- NO / no real info}
The uncertainties on the integrated luminosities for all ZEUS HERA\,II samples
were also re-evaluated using the final understanding of the luminosity 
system~\cite{Zlumi3}. The uncertainty is 1.8\,\% for almost all
data taking periods. The uncorrelated part of this uncertainty is 1\,\%.

\section{Combined QCD and EW analysis}

The analysis presented here was performed at NLO in QCD.
The DGLAP~\cite{Gribov:1972ri,Gribov:1972rt,Lipatov:1974qm,Dokshitzer:1977sg,Altarelli:1977zs}
formalism was used to describe the evolution of the PDFs
with $Q^2$.
The PDFs were parameterised at a starting scale of 1.9\,GeV$^2$.
The analysis followed the method used to extract
the set of PDFs called
HERAPDF2.0~\cite{HERAPDF20} and 
its variants. 
The cross sections as predicted by perturbative QCD
were fitted to the measured cross sections and PDF parameters were determined
through $\chi^2$ minimisation. 
The fits were performed with the ZEUSFitter 
package\,\footnote{The ZEUSFitter package was previously 
used to extract the PDF sets of HERAPDF1.0~\cite{HERAIcombi}
%ZEUS-S and ZEUS-Jets 
and to cross check the HERAPDF2.0 fits.}
and cross-checked with 
the HERAFitter~\cite{HERAFitter} package.

To extract electroweak parameters, either the couplings of the
$Z$ boson to the 
$u$- and $d$-type quarks or 
$\sin^2\theta_W$ and $M_W$
were additional free parameters in the fit. 
The resulting PDFs are called ZEUS-EW.

The PDG14~\cite{PDG14} value $M_Z=91.1876$\,GeV was used throughout the
analysis.
The PDG14 on-shell value of $\sin^2\theta_W=0.22333$~\cite{PDG14} and
the corresponding SM couplings of the $Z$ boson 
to $u$- and $d$-type quarks
were used unless these quantities were free parameters in the fits.
The vector couplings of the $Z$ boson to electrons
were calculated with the PDG14 on-shell value of $\sin^2\theta_W$ 
and kept fixed throughout the analysis unless $\sin^2\theta_W$ 
was a free parameter.
In that case, all couplings of the $Z$ were recalculated according
to the SM formulas.
The PDG14 value of $M_W=80.385$\,GeV was used
unless $M_W$ was a free parameter.
The PDG14 value of $G_F = 1.1663787 \cdot 10^{-5}$\,GeV$^{-2}$ 
was used unless $M_W$ or $\sin^2\theta_W$ were 
free parameters in the fit, 
see Eq.~\ref{eqn:GF}.

% Degrees of freedom 2942 points -17 =2925 for zeus-ew-z
% And 2942-14=2928 for zeus-ew-w
%number of polarised data points (Brian, line 172-174):
%total: 501
%NC:   360
%CC:   141 
Only data with $Q^2 \ge 3.5$\,GeV$^2$ were considered in the analysis.
This gives 2942 cross-section points, of which 501 (360 NC and 141 CC)
are cross sections
measured by ZEUS for polarised beams.
%\,\footnote{It should be noted that
%501 was the highest confirmed score of a batter ever achieved in a 
%professional cricket match.}.
Detailed information on these ZEUS data on cross sections 
for polarised beams
are given\,\footnote{The data sets were listed
as ZEUS NC and ZEUS CC for HERA\,II $E_p=920~$GeV in Table~1 of a
previous publication~\cite{HERAPDF20}.}
in Table~\ref{tab:pol}. 
The number of cross sections used as input to the analysis 
presented here is much larger than
for HERAPDF2.0, because the data sets from ZEUS and H1 were not combined
and, in addition, polarisation was considered for the ZEUS data sets as
listed in Table~\ref{tab:pol} doubling the cross-section values for
these data sets.

All QCD parameters and 
settings entering the analysis were chosen as for HERAPDF2.0
unless explicitly stated. 
The experimental uncertainty, denoted ``experimental/fit'' in the following,
is the uncertainty determined
by the fit using the Hessian method. 
The model uncertainties were 
computed exactly as for 
HERAPDF2.0, except for  the strange-sea contribution, which was assumed to 
be a fixed fraction of the $d$-type sea.

The PDFs parameterised are the gluon distribution, $xg$, 
and the quark distributions in the general form
\begin{linenomath*}\begin{equation}
 xf(x) = A x^{B} (1-x)^{C} (1 + D x + E x^2)~~.
\label{eqn:pdf}
\end{equation}\end{linenomath*}
The quark distributions are
the valence-quark distributions, $xu_v$, $xd_v$, and 
the $u$-type and $d$-type anti-quark distributions,
$x\bar{U}$, $x\bar{D}$. The relations $x\bar{U} = x\bar{u}$ and 
$x\bar{D} = x\bar{d} +x\bar{s}$ are assumed at the starting scale.
A detailed discussion on this parameterisation $ansatz$ can be found
in the HERAPDF2.0 publication~\cite{HERAPDF20}. 
A slight deviation from the HERAPDF2.0 analysis is the 
reduction from 14 to 13 PDF parameters
as described below. 

The parameterisation of the proton PDFs chosen for ZEUS-EW is
\begin{eqnarray}
\label{eq:xgpar}
xg(x) &=   & A_g x^{B_g} (1-x)^{C_g} - A_g' x^{B_g'} (1-x)^{C_g'}  ,  \\
\label{eq:xuvpar}
xu_v(x) &=  & A_{u_v} x^{B_{u_v}}  (1-x)^{C_{u_v}}\left(1+E_{u_v}x^2 \right) , \\
\label{eq:xdvpar}
xd_v(x) &=  & A_{d_v} x^{B_{d_v}}  (1-x)^{C_{d_v}} , \\
\label{eq:xubarpar}
x\bar{U}(x) &=  & A_{\bar{U}} x^{B_{\bar{U}}} (1-x)^{C_{\bar{U}}} , \\
\label{eq:xdbarpar}
x\bar{D}(x) &= & A_{\bar{D}} x^{B_{\bar{D}}} (1-x)^{C_{\bar{D}}} .
\end{eqnarray}

The normalisation parameters, 
$A_{u_v}, A_{d_v}, A_g$, are constrained 
by the quark-number sum rules and the momentum sum rule. 
The parameters $B_{\bar{U}}$ and $B_{\bar{D}}$ were replaced 
by a single $B$ parameter for the sea distributions. 
The strange-quark distribution is expressed 
as an $x$-independent fraction, $f_s$, of the $d$-type sea, 
$x\bar{s}= 0.4\, x\bar{D}$ at the starting scale.
The parameter ${C_g'}$ is fixed to  ${C_g'} = 25$\cite{Martin:2009iq}.
The reduction to 13 parameters was implemented
by replacing
$x\bar{U}(x) =  A_{\bar{U}} x^{B_{\bar{U}}} (1-x)^{C_{\bar{U}}}\left(1+D_{\bar{U}}x\right)$
used for HERAPDF2.0 with 
Eq.~\ref{eq:xubarpar}.  
The reduction to 13 PDF parameters for ZEUS-EW greatly improved
the stability of the fits necessary to determine the 
parameterisation uncertainties.

A 13-parameter fit with fixed SM $Z$ couplings, $\sin^2\theta_W$ and $M_W$,
called ZEUS-13p, was performed as a reference.
A fit with 13+4 parameters, called ZEUS-EW-Z,
was used to extract the four couplings of the
$Z$ to $u$- and $d$-type quarks.
Two (13+1)-parameter fits 
called ZEUS-EW-S and ZEUS-EW-W were
used to extract $\sin^2\theta_W$ and $M_W$ separately,
while keeping the other one fixed.
In addition, a (13+2)-parameter fit 
called ZEUS-EW-S-W  was performed to extract  simultaneously 
$\sin^2\theta_W$ and $M_W$.
As  cross-checks, fits in which the PDF parameters were fixed to 
ZEUS-13p and only the electroweak parameters were allowed 
to vary were also performed. 

The parameterisation uncertainties for all fits were obtained by adding
extra $D$ and $E$ parameters one by one to the fit. 
It was checked whether this
caused  a significant change in the result on the EW parameters.
It turned out that  
only adding back the parameter $D_{\bar{U}}$ or adding the parameter $D_g$
resulted in significant differences.
If a (14+4)-parameter fit including the
parameter $D_{\bar{U}}$ would have been chosen for ZEUS-EW, 
the determination of the parameterisation uncertainties 
would have required (15+4)-parameter fits. Such fits were found
to be too unstable to provide reliable uncertainties.

The parameter $D_{\bar{U}}$ was added for the extraction
of HERAPDF2.0 because it reduced the overall $\chi^2$ by
about 0.005 per degree of freedom to 1357/1131=1.200.
The  $\chi^2$ per degree of freedom of ZEUS-13p is
$3275/2929 =1.118$. If $D_{\bar{U}}$ would have been added as a 14th
parameter, a reduction of $\chi^2$ similar to the reduction for
HERAPDF2.0 would have been obtained.
However, the instability of the
(15+4)-parameter fits was considered to outweigh
this minimal gain in  $\chi^2$. 
The $\chi^2/{\rm dof}$ values of all ZEUS-EW fits are similar
to the values for ZEUS-13p.
As 14-parameter fits were used to evaluate the parameterisation
uncertainties, the uncertainties associated with the $D_{\bar{U}}$
are included.
 
All results were cross-checked with fits at NNLO QCD, which yielded
compatible results.  
However, as the EW analysis is partially at LO, see below, 
a treatment of the PDFs at NLO was considered more consistent, 
because $\alpha_s^2$ is of the same order of magnitude as $\alpha$.

The uncertainties on the polarisation 
as listed in Table~\ref{tab:pol} were taken into account
in all fits presented in this paper.
However, it was found that the effect of these uncertainties
is negligible compared to the total experimental/fit uncertainty.

As described in Section~\ref{sec-pred}, the reduced cross sections 
used as input to the analysis were published 
by the individual collaborations after QED corrections
were applied.
These corrections are mostly on the
percent level, but reach 15\,\% for a few cross sections.
The correction
factors were calculated by producing 
Monte Carlo data sets 
for which radiative corrections were
either turned on or off for comparison. 
This was done with the program 
{\sc Heracles}~\cite{cpc:69:155} interfaced to the
hadronisation programs within the program 
{\sc Djangoh}\cite{proc:hera:1991:1419}.
However, the two collaborations did not use 
the {\sc Heracles} program with exactly the same options.
The ZEUS collaboration only corrected for LO initial-
and final-state radiation of the electron.
The H1 collaboration included the effects of quark radiation
and $Z$ self-energy~\cite{H1allhQ2}~\footnote{The term $Z$ self-energy
denotes the influence of vacuum polarisation~\cite{Spiesberger:95}.}.
The difference introduced by these extra contributions
is, however, always less than 1\,\%~\cite{Bardin:89}.
The H1 collaboration published~\cite{H1allhQ2} a cross-check with the
programs {\sc Hector}~\cite{Hector} and EPRC~\cite{Spiesberger:95} and 
concluded that the uncertainties are below 2\,\% in all of the phase space.
In addition, the effect of the exchange of two or more photons between the
electron and the quarks, which was not implemented 
in {\sc Heracles}, was found to 
be negligible. 
The H1 collaboration included phase-space-dependent 
uncertainties in the uncorrelated uncertainties of their 
published cross sections.
The ZEUS collaboration did not assign any uncertainties to their
QED corrections. 
%The LO initial- and final-state corrections
%were assumed to be precisely known.
As a cross-check, an extra uncertainty of the size assigned by H1
was also added to the uncorrelated uncertainties
on the ZEUS cross sections
for polarised beams.
In all cases, the effect on the extracted EW parameters was negligible.

The published cross sections were not corrected for 
further electroweak effects by either ZEUS or H1.
For the analysis presented here, electroweak effects
were taken into account through $\Delta R$ as introduced
in Eq.~\ref{eq:rewrite1}. It was 
computed with the program EPRC~\cite{Spiesberger:95},
where weak box-diagrams, $\gamma/Z$ interference and
$Z$ and $W$ self-energies were taken into account.   
The running of $\alpha$, relevant for the CC cross sections,
is also absorbed in $\Delta R$.

\section{Couplings of the \boldmath{$Z$} boson 
         to the \boldmath{$u$} and \boldmath{$d$} quarks}

To determine the axial-vector and vector couplings of the $Z$ to the
$u$- and $d$-type quarks, 
$a_u, v_u, a_d, v_d$,
the QCD predictions depending  
on the 13 PDF parameters plus the four couplings
were fitted simultaneously to the data.
The fit as well as the resulting set of PDFs are called ZEUS-EW-Z.

\begin{figure}[p]
\vfill
\begin{center}
\includegraphics[width=6in]{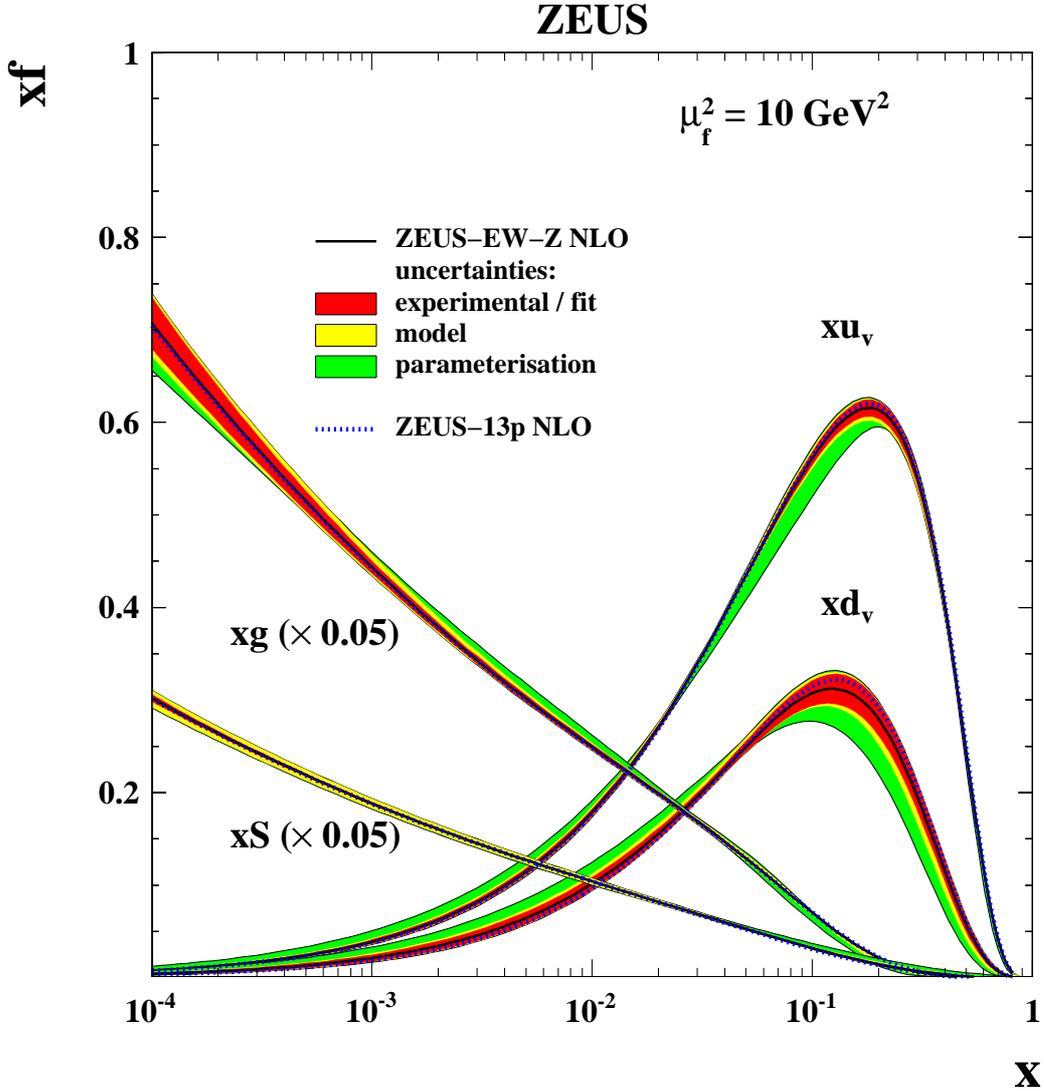}
\end{center}
\caption{The PDF set ZEUS-EW-Z with
         cumulative  
         experimental/fit, model and
         parameterisation uncertainties at the 
         factorisation scale $\mu_{\rm f}^2=10$\,GeV$\,^2$.
         All positive and negative model uncertainties 
         were added 
         separately in quadrature.
         The parameterisation uncertainty represents
         an envelope of all individual parameterisation
         uncertainties.  
         Also shown are the 
         central values of the reference fit ZEUS-13p.         
}
\label{fig:ZEUS-EW-Z-13p}
\vfill
\end{figure}

\begin{figure}[p]
\vfill
\begin{center}
\includegraphics[width=6in]{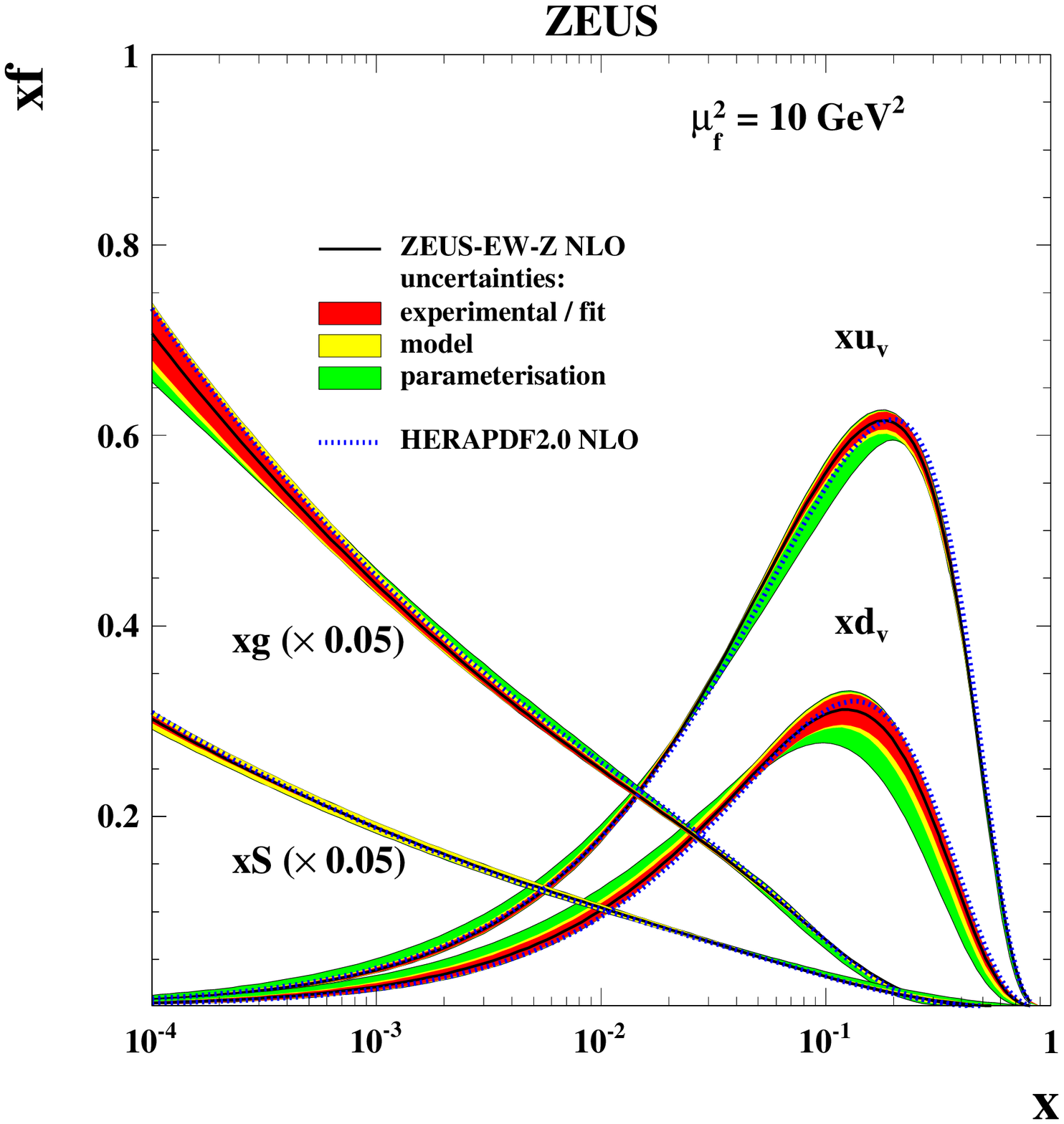}
\end{center}
\caption{The PDF set ZEUS-EW-Z with
         cumulative  
         experimental/fit, model and
         parameterisation uncertainties at the 
         factorisation scale $\mu_{\rm f}^2=10$\,GeV$\,^2$.
         Also shown are the 
         central values of HERAPDF2.0 NLO.
         Other details as in Fig.~\protect \ref{fig:ZEUS-EW-Z-13p}.
}
\label{fig:ZEUS-EW-Z-HPDF}
\vfill
\end{figure}

A comparison of the PDFs of ZEUS-EW-Z with full uncertainties to the 
central values of the PDFs of ZEUS-13p is shown
in Fig.~\ref{fig:ZEUS-EW-Z-13p}. 
Within uncertainties, the PDFs of ZEUS-EW-Z agree well
with ZEUS-13p.
The freeing of the couplings in the fit has very little influence on the
PDF parameters.
The full correlation matrix 
is given as Table~\ref{tab:matrix}.
The small correlation between PDF parameters
and couplings is a sign that the PDFs are not
absorbing any significant non-SM effects which could show up in 
the electroweak couplings.
A comparison of the PDFs of ZEUS-EW-Z 
to the PDFs of HERAPDF2.0 is shown
in Fig.~\ref{fig:ZEUS-EW-Z-HPDF}. 
The PDFs agree well within uncertainties.

\begin{figure}[p]
\vfill
\begin{center}
\includegraphics[width=6in]{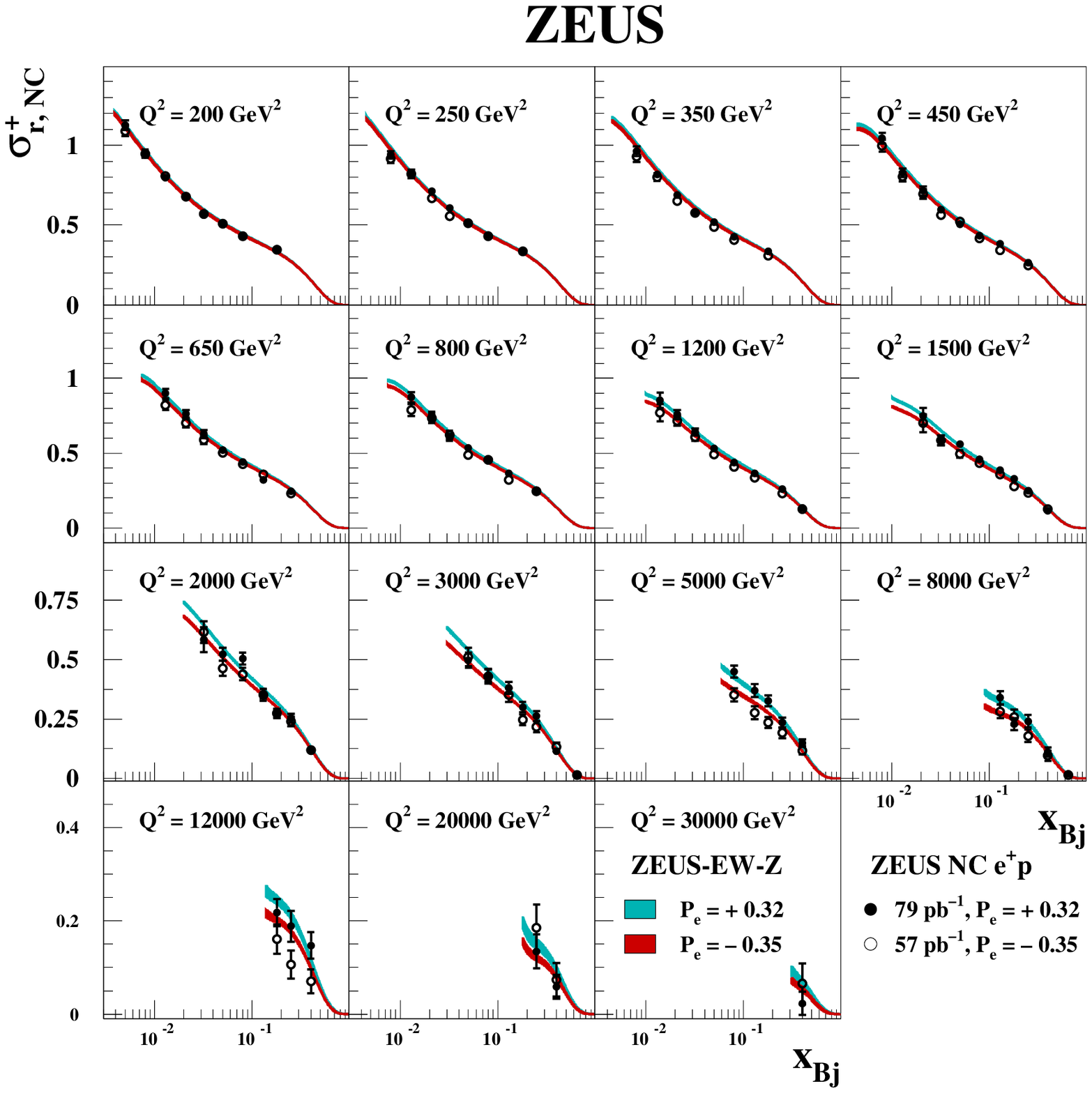}
\end{center}
\caption{The predictions of ZEUS-EW-Z compared to 
  the $e^+p$ NC DIS reduced cross-section $\sigma^+_{r,NC}$
  for positively and negatively polarised beams
  plotted as a function of $x$ at fixed $Q^2$.
  The closed (open) circles represent the ZEUS data
  for positive (negative) polarisation.
  The bands indicate the full uncertainty on the 
  predictions of ZEUS-EW-Z.
}
\label{fig:NCpp}
\vfill
\end{figure}

\begin{figure}[p]
\vfill
\begin{center}
\includegraphics[width=6in]{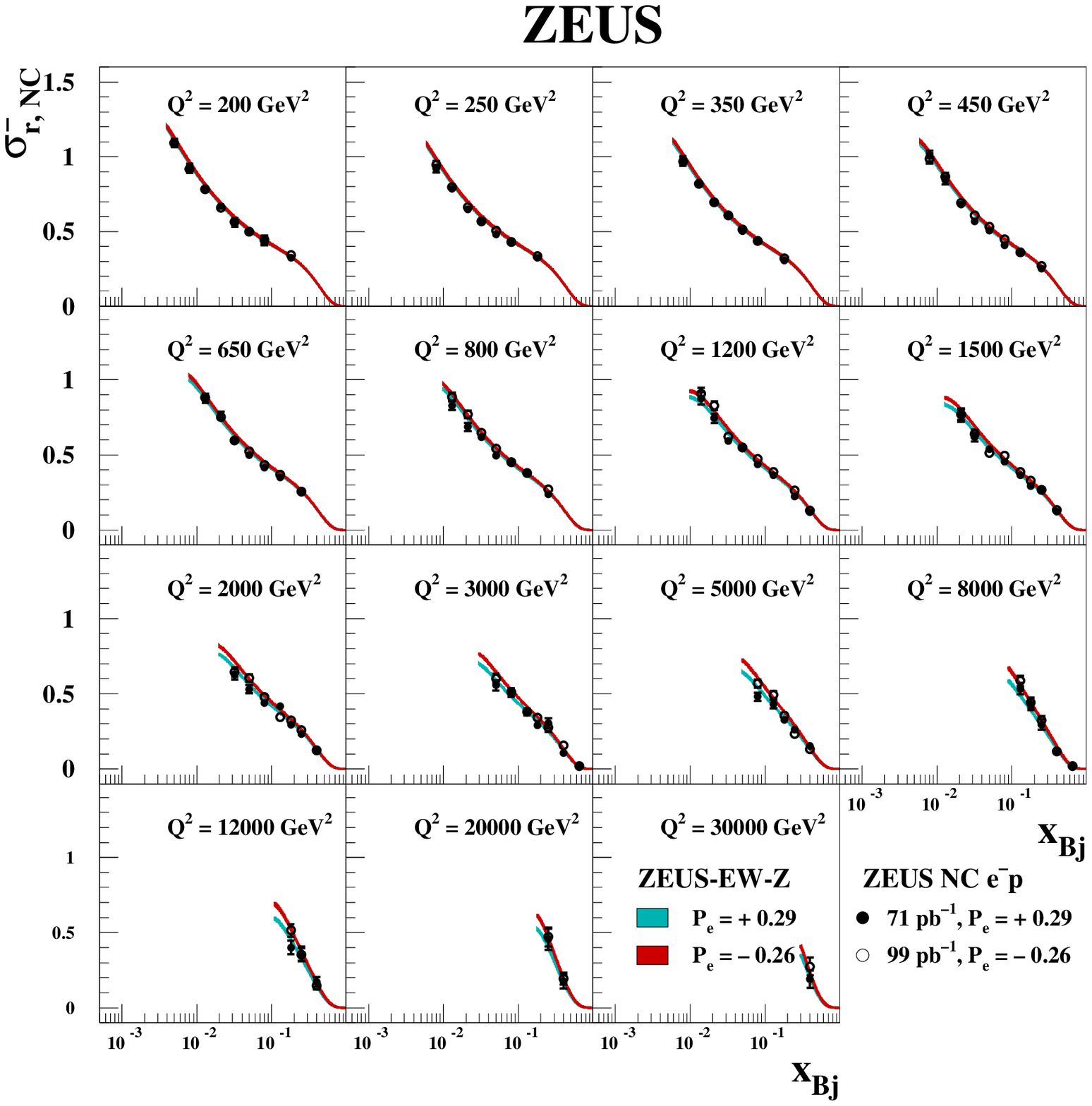}
\end{center}
\caption{The predictions of ZEUS-EW-Z compared to 
  the $e^-p$ NC DIS reduced cross-section $\sigma^-_{r,NC}$
  for positively and negatively polarised beams
  plotted as a function of $x$ at fixed $Q^2$.
  The closed (open) circles represent the ZEUS data
  for positive (negative) polarisation.  
  The bands indicate the full uncertainty on the 
  predictions of ZEUS-EW-Z.
}
\label{fig:NCep}
\vfill
\end{figure}

The predictions of ZEUS-EW-Z are compared to the ZEUS reduced NC cross sections
in Figs.~\ref{fig:NCpp} and~\ref{fig:NCep} for $e^+p$ and $e^-p$ scattering,
respectively. In both cases, data with positive and negative
beam polarisation are shown separately. ZEUS-EW-Z describes the data 
well.   

The values of the couplings were determined in the simultaneous fit as
\begin{eqnarray}
\nonumber
a_u & = & +0.50~ ^{+0.09}_{-0.05} \,{\rm {\scriptstyle{(experimental/fit)}}}~ 
        ^{+0.04}_{-0.02}\,{\rm {\scriptstyle{(model)}}}~ 
        ^{+0.08}_{-0.01}\,{\rm {\scriptstyle{(parameterisation)}}} ~,\\
\nonumber
a_d & = & -0.56~ ^{+0.34}_{-0.14}\,{\rm {\scriptstyle{(experimental/fit)}}}~ 
        ^{+0.11}_{-0.05}\,{\rm {\scriptstyle{(model)}}}~ 
        ^{+0.20}_{-0.00}\,{\rm {\scriptstyle{(parameterisation)}}} ~,\\
\nonumber
v_u & = & +0.14~ ^{+0.08}_{-0.08}\,{\rm {\scriptstyle{(experimental/fit)}}}~ 
        ^{+0.01}_{-0.02}\,{\rm {\scriptstyle{(model)}}}~
        ^{+0.00}_{-0.03}\,{\rm {\scriptstyle{(parameterisation)}}}  ~,\\
\nonumber
v_d & = & -0.41~ ^{+0.24}_{-0.16}\,{\rm {\scriptstyle{(experimental/fit)}}}~ 
        ^{+0.04}_{-0.07}\,{\rm {\scriptstyle{(model)}}}~ 
        ^{+0.00}_{-0.08}\,{\rm {\scriptstyle{(parameterisation)}}} ~.
\end{eqnarray}

They are also listed in Table~\ref{tab:Z} with their experimental/fit and 
total uncertainties
and compared to  SM predictions. 
Also listed are values obtained
in a fit where the only free parameters were the $Z$ couplings and the PDFs
were fixed to ZEUS-13p. These values for the couplings are compatible to 
those obtained by ZEUS-EW-Z. 
This cross-check confirms that
the determination of the $Z$ couplings is essentially decoupled from
the QCD part of the fit.

Another fit, HPDF1, was performed with the $Z$ couplings free and 
the PDFs fixed to HERAPDF2.0 NLO. 
The results, also listed in Table~\ref{tab:Z}, are
in agreement with ZEUS-EW-Z. 
It should be noted that HERAPDF2.0 was extracted using
a different value of $\sin^2\theta_W$.
Therefore, a fit HPDF2 
using this $\sin^2\theta_W$ value was also performed.
The result is also listed in Table~\ref{tab:Z}.
The values agree well within uncertainties with those from
the fit using the on-shell value.

\begin{figure}[p]
\vfill
\begin{center}
\includegraphics[width=4in]{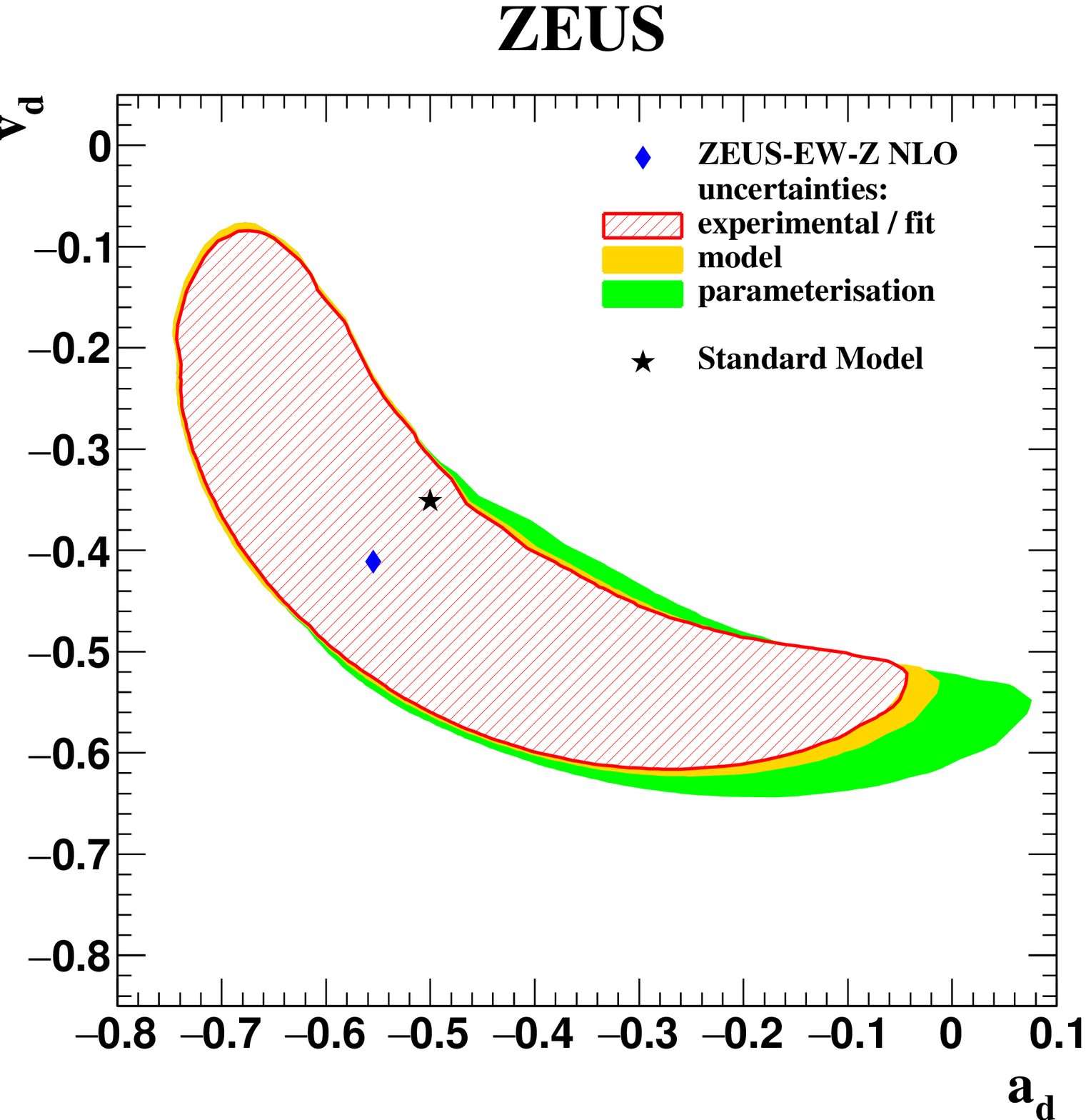}
\includegraphics[width=4in]{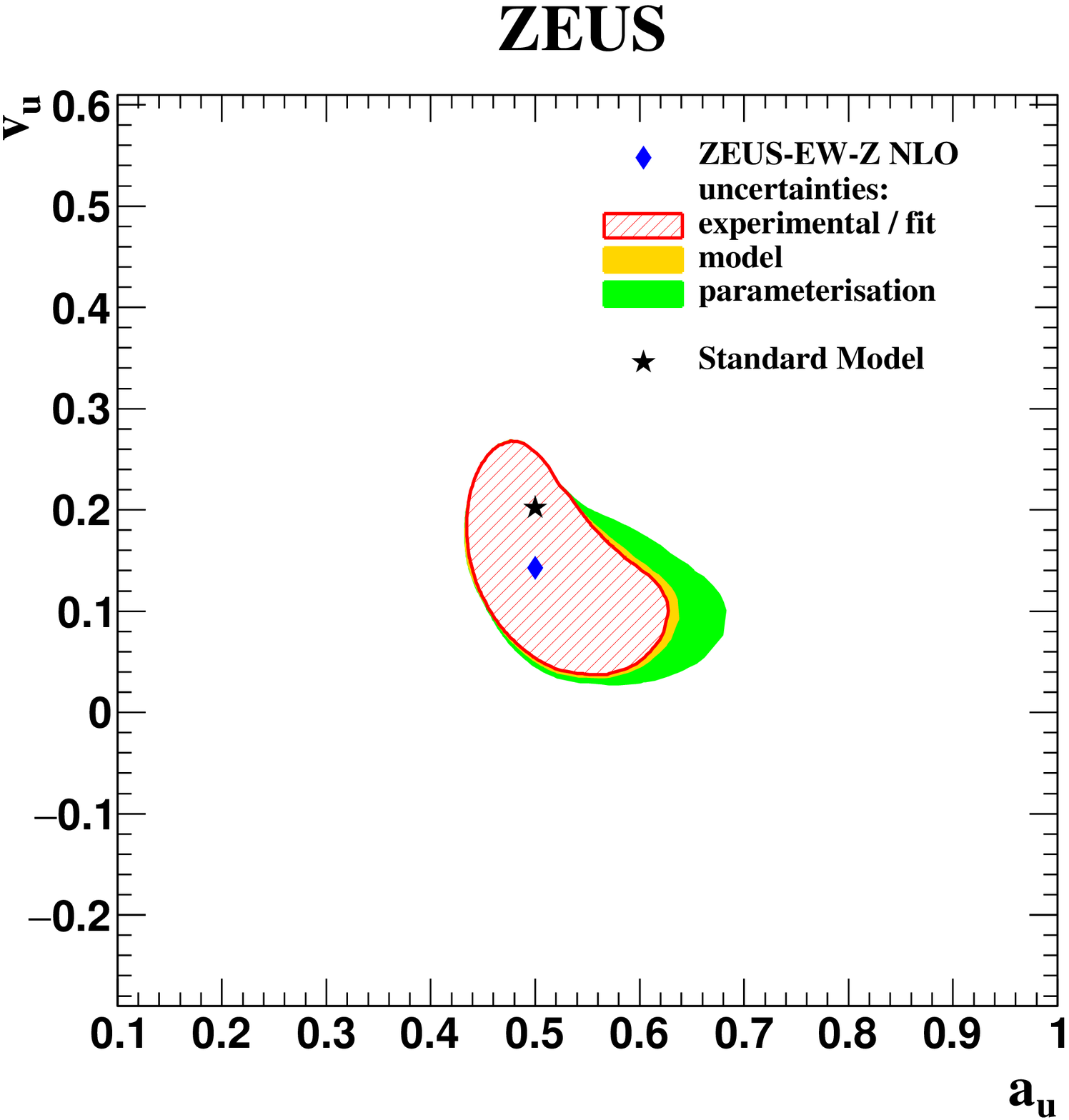}
\end{center}
\caption{The 68\,\%\,C.L. contours 
         for ($a_d,v_d$) and ($a_u,v_u$)
         obtained for the ZEUS-EW-Z fit.
}
\label{fig:av-us}
\vfill
\end{figure}

\begin{figure}[p]
\vfill
\begin{center}
\includegraphics[width=4in]{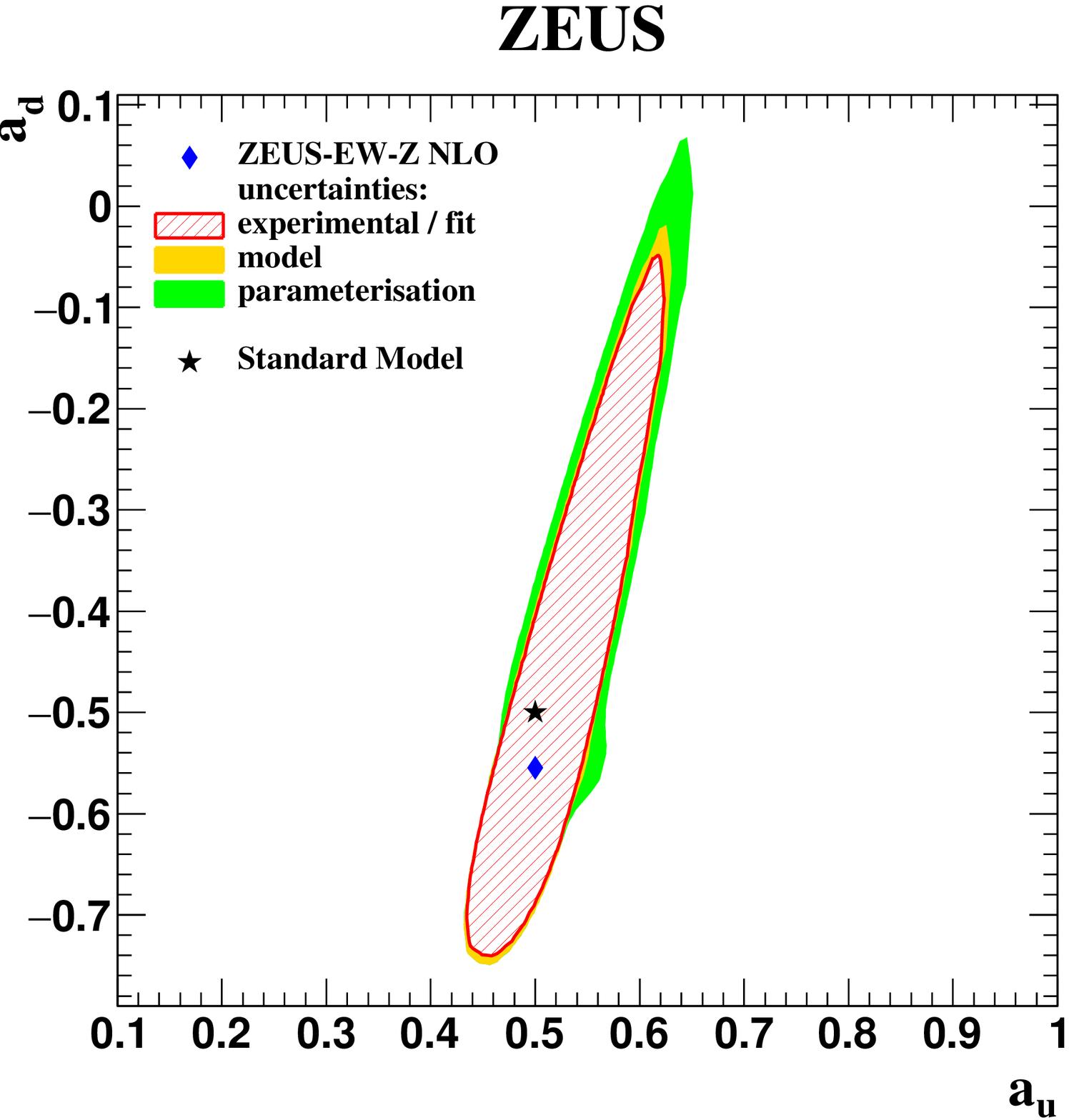}
\includegraphics[width=4in]{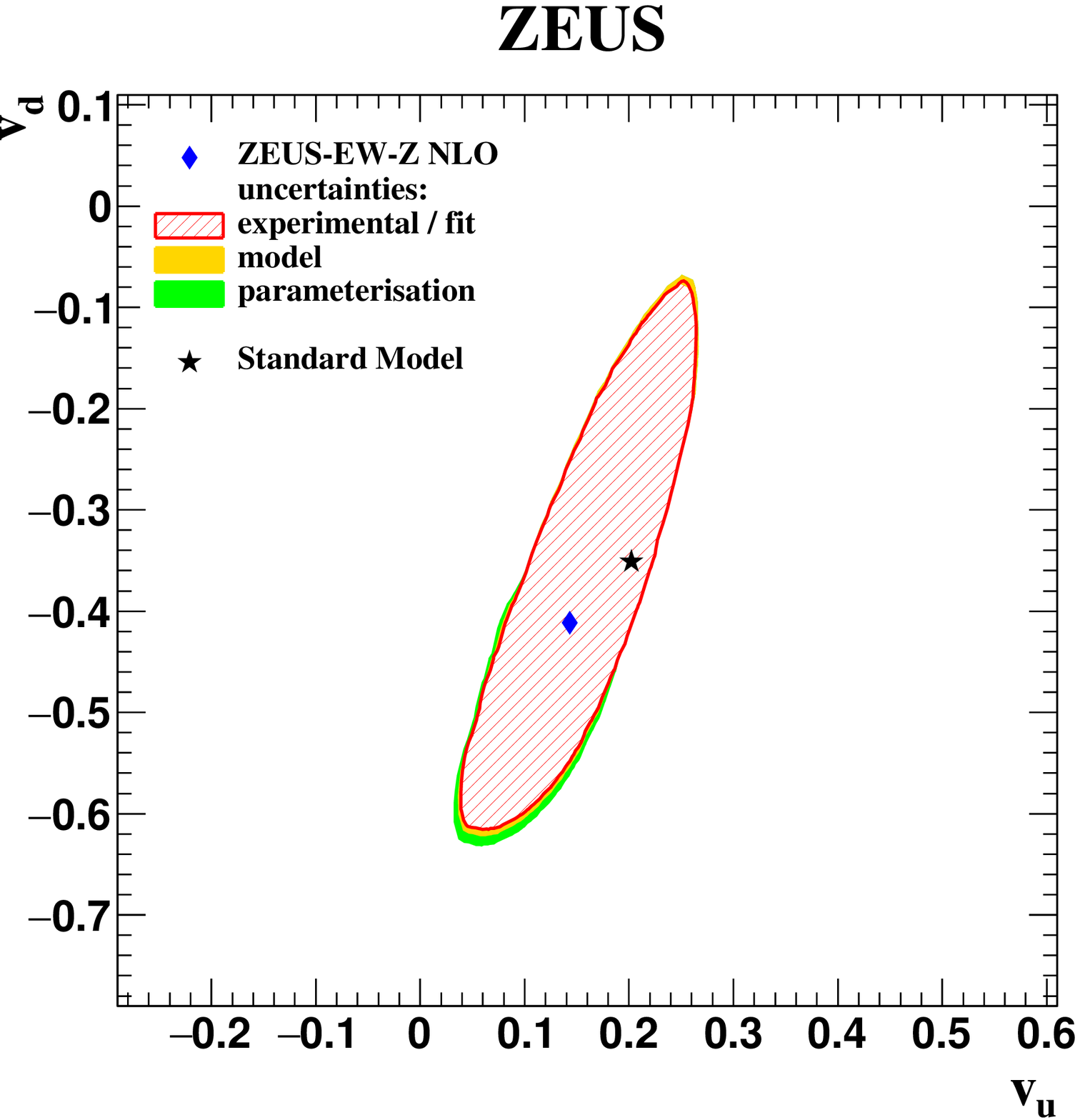}
\end{center}
\caption{The 68\,\%\,C.L. contours 
         for ($a_u,a_d$) and ($v_u,v_d$)
         obtained for the ZEUS-EW-Z fit.
}
\label{fig:aavv}
\vfill
\end{figure}

The correlations between the four couplings 
obtained in ZEUS-EW-Z are listed as 
part of Table~\ref{tab:matrix}.
Two-dimensional scans 
were performed to obtain 
so-called profile likelihood contours. 
The two parameters under investigation were modified in small steps.
For each point, a fit was performed to minimise $\chi^2$
with respect to all other parameters. The $\chi^2$-values thus calculated 
were used to obtain 68\,\%\,C.L. contours.
The results for the couplings $a_u,v_u$ and $a_d,v_d$ are
shown in Fig.~\ref{fig:av-us}~\footnote{Numerical information 
is available as additional material for this publication.}.
Figure~\ref{fig:aavv} shows the 68\,\%\,C.L. 
contour plots for $a_u,a_d$ and $v_u,v_d$. 
This illustration demonstrates very clearly that
the HERA data constrain the couplings of the $Z$ boson to 
the $u$ quark significantly better
than the couplings to the $d$ quark.
All measurements are compatible with the SM.
The parameterisation uncertainties mostly arise from
the $D_{\bar{U}}$ parameter. As this parameter is constrained to be positive,
the axial-vector couplings can only increase due to these uncertainties.

\begin{figure}
\vfill
\begin{center}
\includegraphics[width=4in]{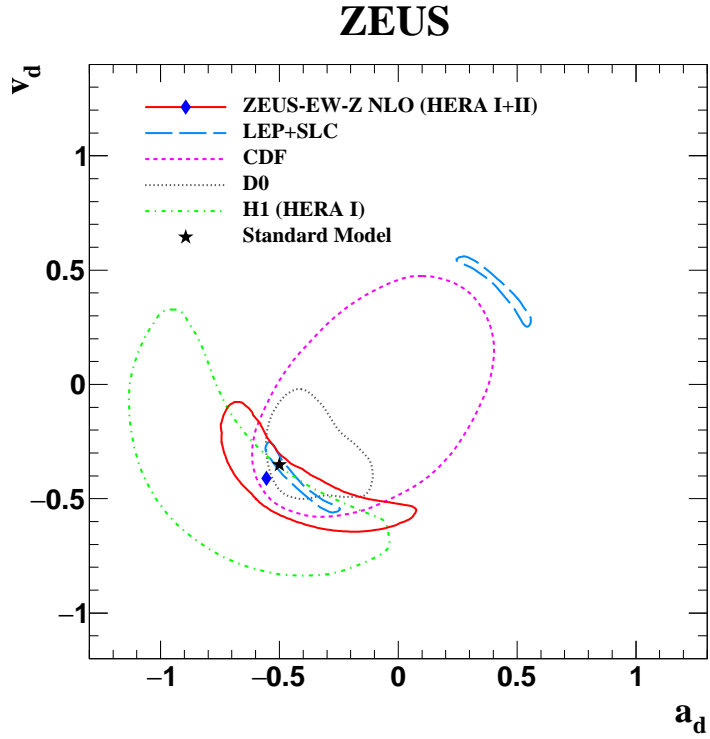}
\includegraphics[width=4in]{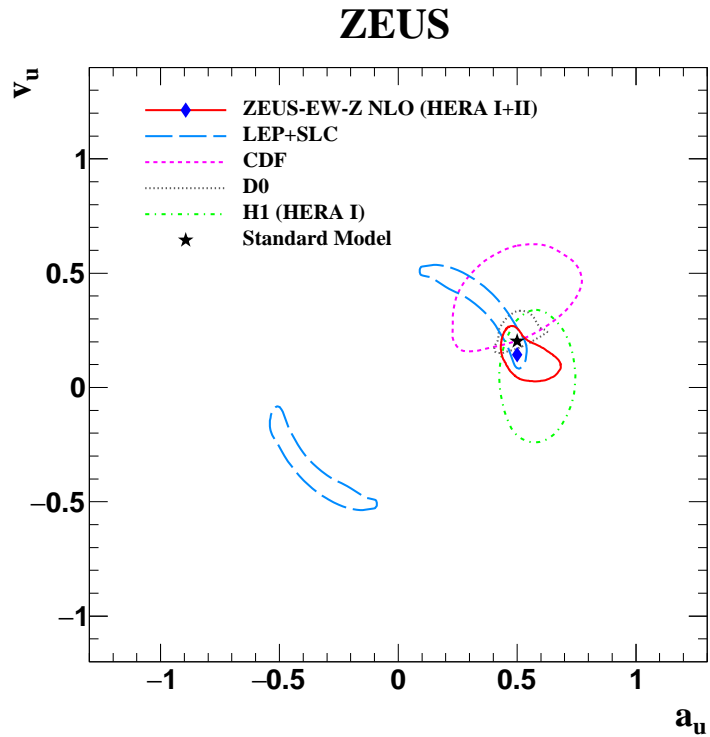}
\end{center}
\caption{The 68\,\%\,C.L. contours 
         for ($a_d,v_d$) and ($a_u,v_u$)
         obtained for the ZEUS-EW-Z fit.
         Also shown are  
         results from LEP (ALEPH, OPAL, L3 and DELPHI) 
         plus SLC (SLD) combined, %~\cite{ZLEPSLC}, 
         the Tevatron (CDF and D0), %~\cite{ZD0Old,ZCDF} 
         and HERA\,I (H1).%~\cite{ZH1}.
}
\label{fig:av-all}
\vfill
\end{figure}

\begin{figure}[p]
\vfill
\begin{center}
\includegraphics[width=6in]{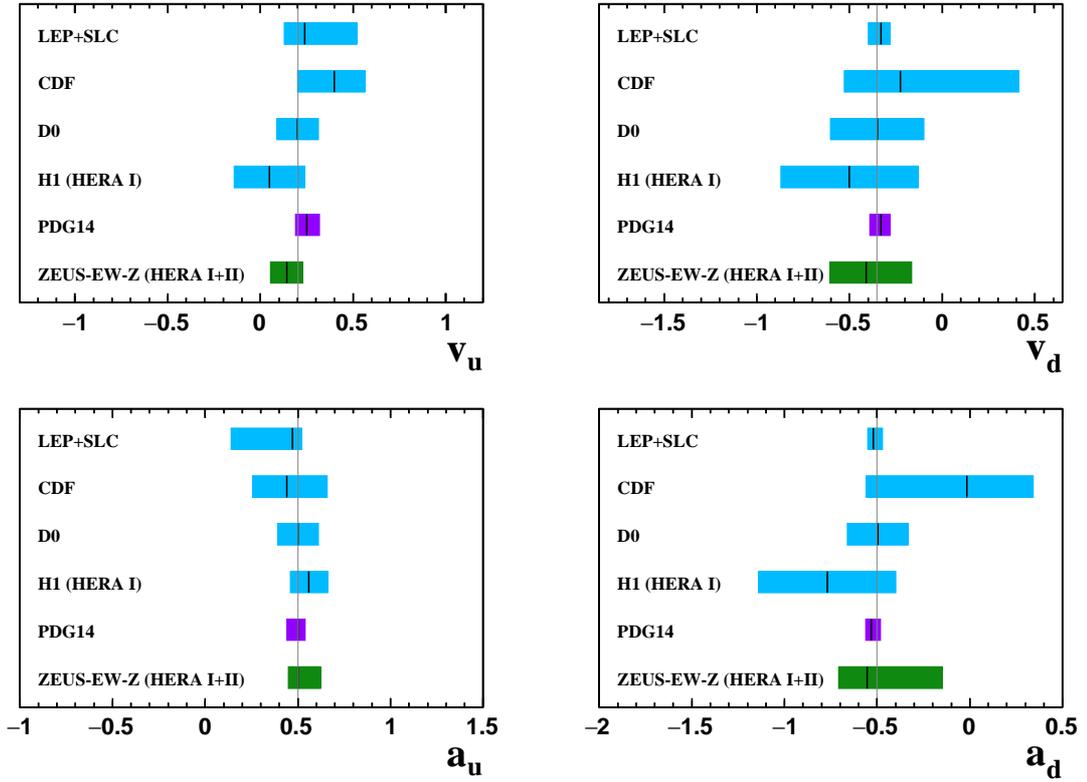}
\end{center}
\caption{The values obtained by the ZEUS-EW-Z fit
         for $a_d, a_u, v_d$ and $v_u$ compared to
         results from LEP (ALEPH, OPAL, L3 and DELPHI) 
         plus SLC (SLD) combined, %~\cite{ZLEPSLC}, 
         the Tevatron (CDF and D0), %~\cite{ZD0Old,ZCDF} 
         and HERA\,I (H1) %~\cite{ZH1} 
         and the world average from these individual
         measurements as given by PDG14.   
         Vertical black lines in each box 
         indicate central values,
         the long gray vertical lines indicate the SM
         predictions. The ZEUS-EW-Z result is given with
         total uncertainties.
}
\label{fig:comparison}
\vfill
\end{figure}

The results from ZEUS-EW-Z are compared
to other measurements from LEP+SLC~\cite{ZLEPSLC}, 
the Tevatron~\cite{ZD0Old,ZCDF} and from HERA\,I (H1)~\cite{ZH1}
in Figs.~\ref{fig:av-all} and~\ref{fig:comparison}.
The PDG14 value obtained from these measurements 
is also given in Fig.~\ref{fig:comparison}.
The ZEUS results on the axial-vector 
and vector couplings to $u$-type quarks 
are the most precise published single values.

\section{Electroweak mixing angle and mass of the \boldmath{$W$} boson}

The SM cross sections depend on $\sin^2\theta_W$  through three mechanisms:
\begin{enumerate}
 \item 
  through $\chi_z$, see Eq.~\ref{eqn:chiz};
 \item 
  through the normalisation factor 
  from Eqs.~\ref{eqn:xsec_CC1} and~\ref{eqn:xsec_CC2}
%  through the $W$ propagator term 
  with $G_F$ rewritten as described in Eq.~\ref{eqn:GF};
 \item 
  through the vector couplings of the $Z$ to the quarks. 
\end{enumerate}

The (13+1)-parameter fit ZEUS-EW-S with $M_W$ fixed to the PDG14 value
exploits all three dependencies.
It yields a value for the on-shell 
$\sin^2\theta_W$ of

\begin{linenomath*}\begin{equation}
\nonumber 
  \sin^2\theta_W = 0.2252~ \pm 0.0011 \,{\rm {\scriptstyle{(experimental/fit)}}}~ 
                             ^{+0.0003}_{-0.0001} \,{\rm {\scriptstyle{(model)}}} 
                           ~ ^{+0.0007}_{-0.0001} \,{\rm {\scriptstyle{(parameterisation)}}}~.  
\end{equation}\end{linenomath*}

The world average in PDG14 for the on-shell value is 
$\sin^2\theta_W = 0.22333~ \pm 0.00011 {\rm (total)}$.
The measurement presented here is slightly high 
in comparison to the world average. 
The precision of this result is comparable to the 
experimental precision achieved in the neutrino sector~\cite{NuTeV,PDG14}.
The advantage of the present extraction
is that the nuclear effects that have to be taken into account
in the analysis of neutrino heavy-target data are not present in $ep$ data.
A cross-check was performed with the PDF parameters fixed to
ZEUS-EW-13p. The result 
is $\sin^2\theta_W = 0.2241~ \pm 0.0009\,{\rm {(experimental/fit)}}$,
which is compatible with the result from ZEUS-EW-S.

The three mechanisms as listed above influence the 
result to different degrees.
The first mechanism exploits the NC data.
The influence of the second
mechanism was tested by fixing $G_F$ to its PDG14 value
in Eqs.~\ref{eqn:xsec_CC1} and~\ref{eqn:xsec_CC2}. 
This removed the influence of the CC data and resulted
in an increase of the experimental/fit uncertainty by a factor of three.
This demonstrates that both NC and CC data contribute significantly to the
full precision.
The influence of the third 
mechanism was found to be negligible by fixing the couplings to 
their SM values.

\begin{figure}[p]
\vfill
\begin{center}
\includegraphics[width=6in]{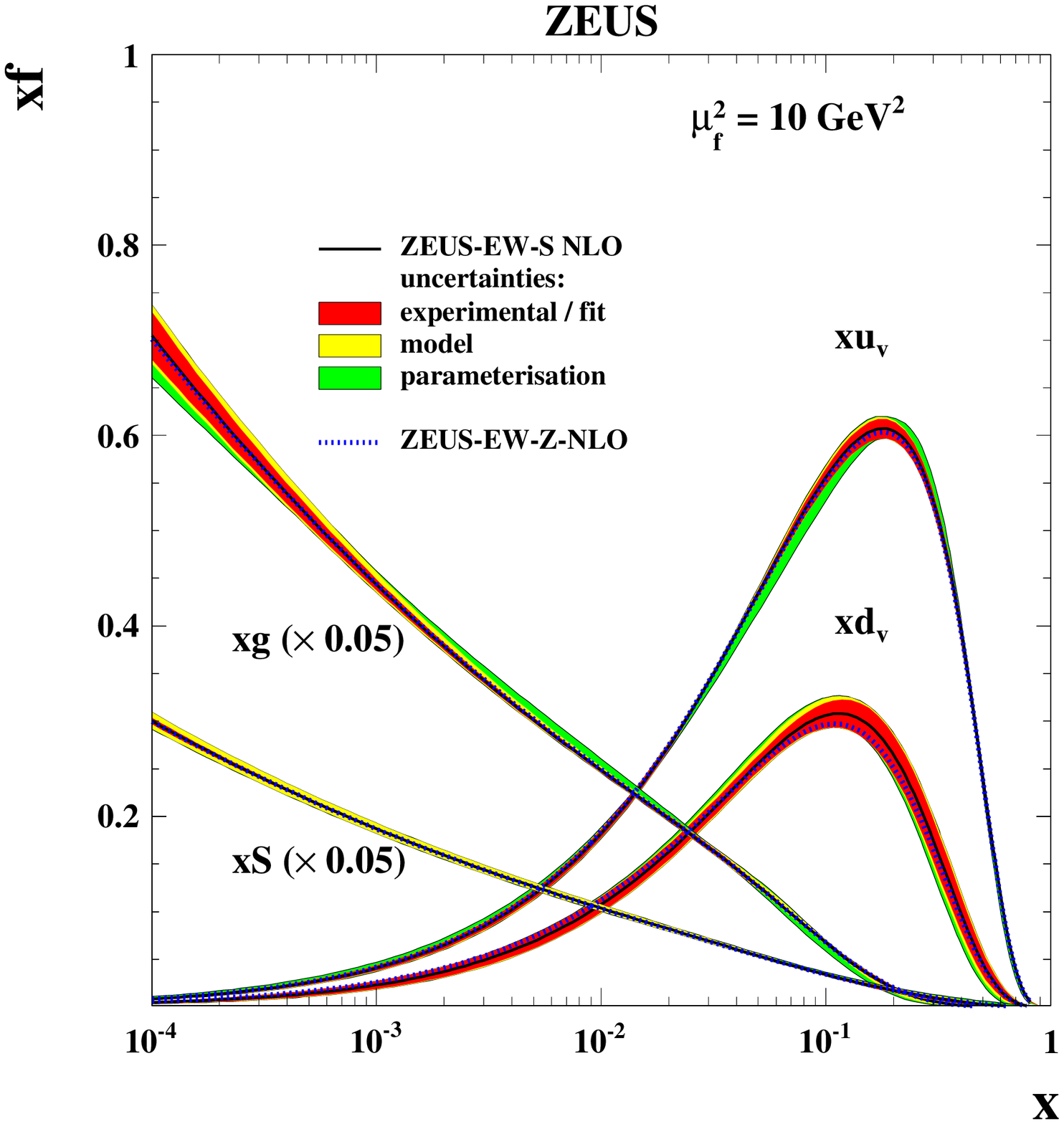}
\end{center}
\caption{The PDF set ZEUS-EW-S with 
         cumulative  
         experimental/fit, model and
         parameterisation uncertainties at the 
         factorisation scale $\mu_{\rm f}^2=10$\,GeV$\,^2$.
         Also shown are the central values  
         of ZEUS-EW-Z.
         Other details as in Fig.~\protect \ref{fig:ZEUS-EW-Z-13p}.
}
\label{fig:ZEUS-EW-S}
\vfill
\end{figure}

\begin{figure}[p]
\vfill
\begin{center}
\includegraphics[width=6in]{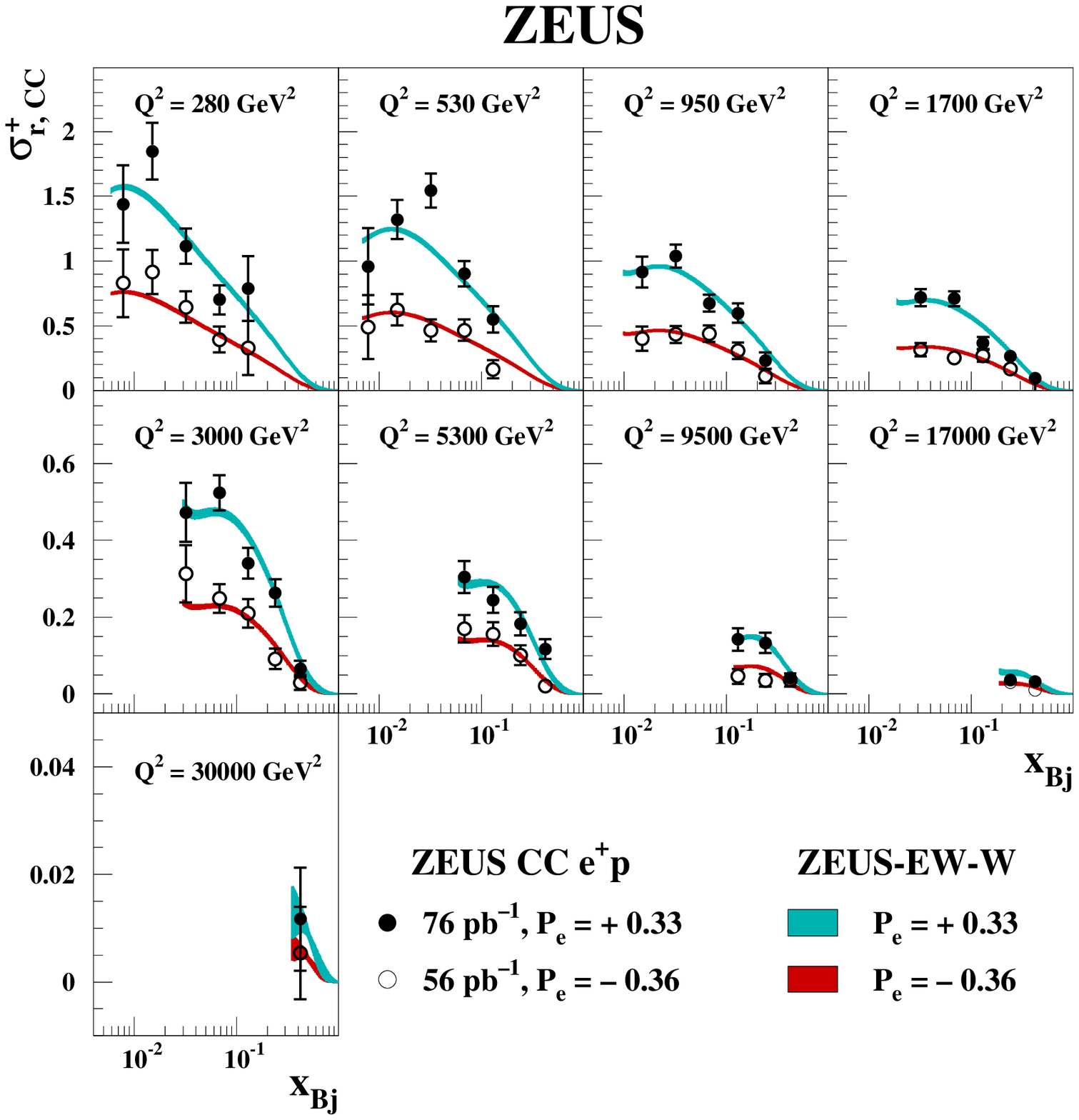}
\end{center}
\caption{The predictions of ZEUS-EW-S compared to 
  the $e^+p$ CC DIS reduced cross-section $\sigma^+_{r,CC}$
  for positively and negatively polarised beams
  plotted as a function of $x$ at fixed $Q^2$.
  The closed (open) circles represent the ZEUS data
  for positive (negative) polarisation.
  The bands indicate the full uncertainty on the 
  predictions of ZEUS-EW-S.
}
\label{fig:CCpp}
\vfill
\end{figure}

\begin{figure}[p]
\vfill
\begin{center}
\includegraphics[width=6in]{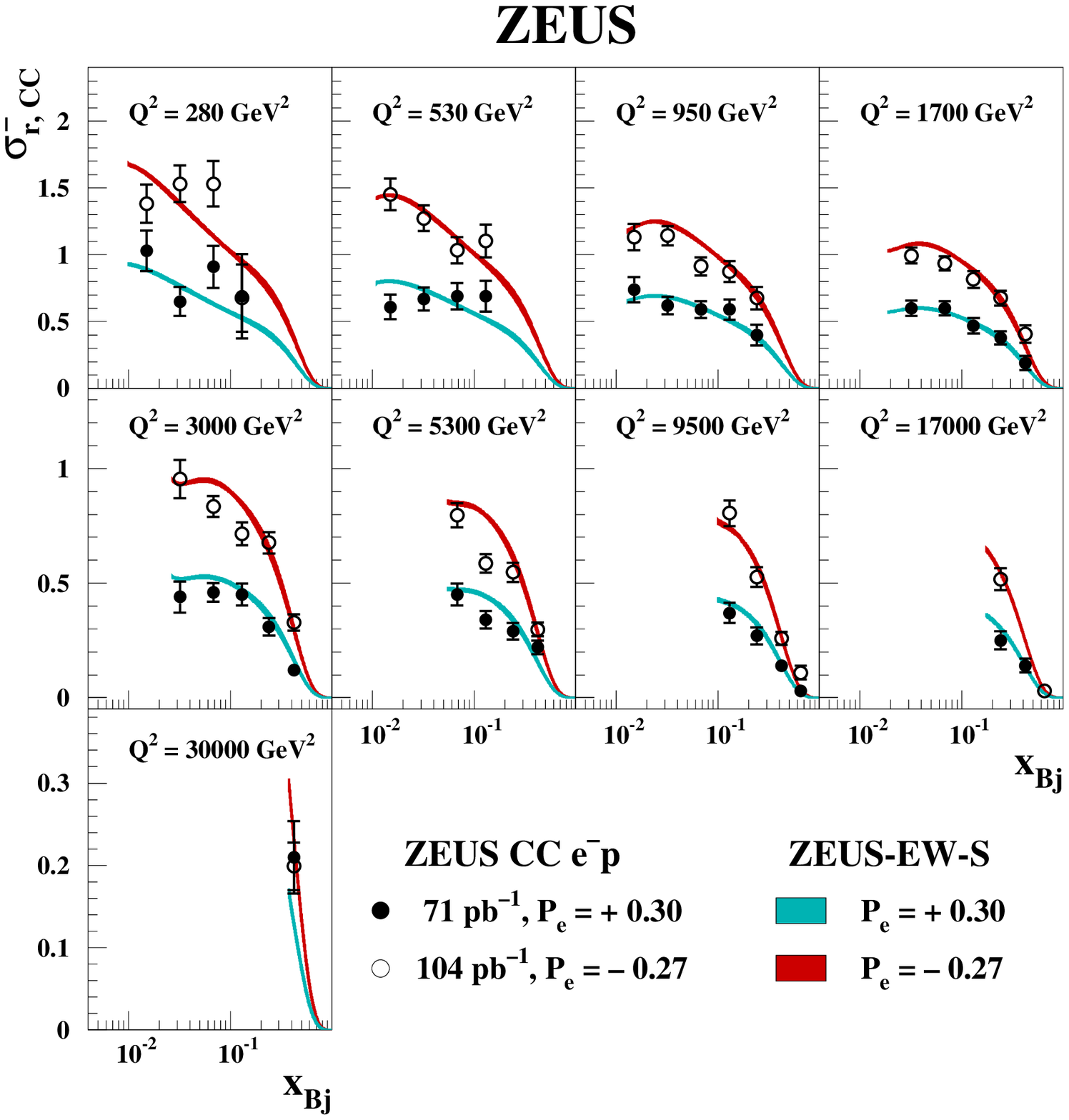}
\end{center}
\caption{The predictions of ZEUS-EW-S compared to 
  the $e^-p$ CC DIS reduced cross-section $\sigma^-_{r,CC}$
  for positively and negatively polarised beams
  plotted as a function of $x$ at fixed $Q^2$.
  The closed (open) circles represent the ZEUS data
  for positive (negative) polarisation.
  The bands indicate the full uncertainty on the 
  predictions of ZEUS-EW-S.
}
\label{fig:CCep}
\vfill
\end{figure}

The PDFs of ZEUS-EW-S are compared to
the PDFs of ZEUS-EW-Z
in Fig.~\ref{fig:ZEUS-EW-S}.
The two sets of PDFs agree very well.
The predictions of ZEUS-EW-S are compared to the reduced CC cross sections
in Figs.~\ref{fig:CCpp} and~\ref{fig:CCep} for $e^+p$ and $e^-p$ scattering,
respectively. In both cases, data with positive and negative
beam polarisation are shown separately; ZEUS-EW-S describes the data 
well.

\begin{figure}[p]
\vfill
\begin{center}
\includegraphics[width=6in]{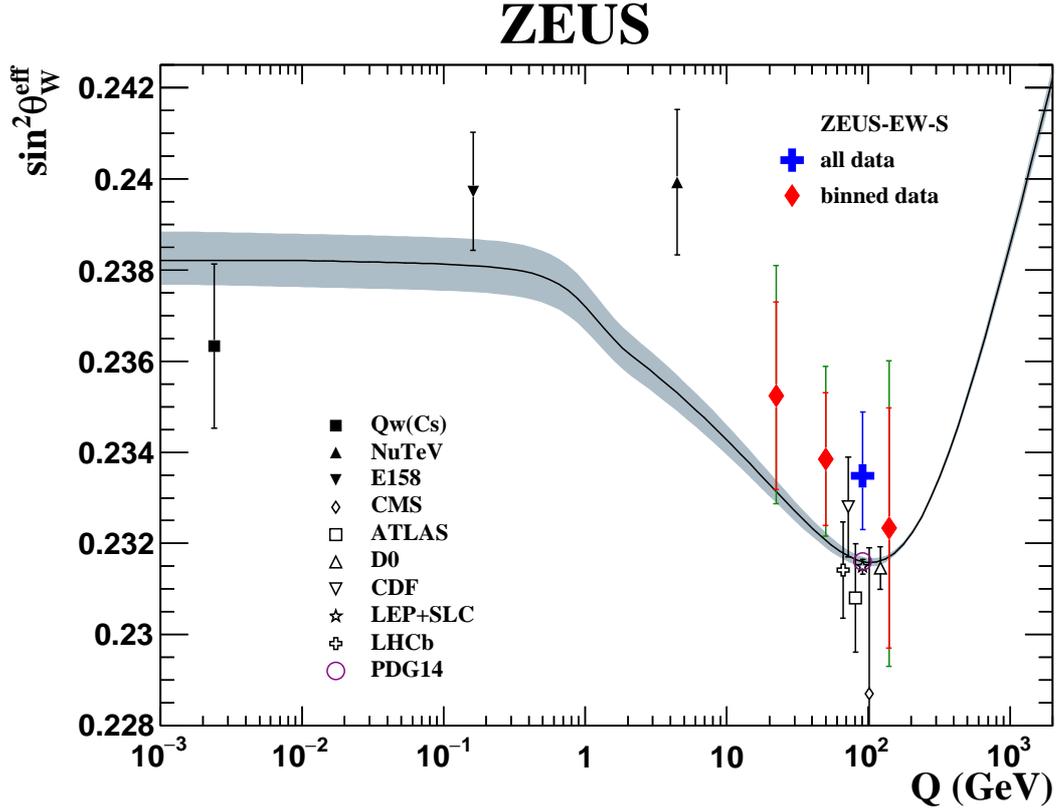}
\end{center}
\caption{The scale dependence of $\sin^2\theta_W^{\rm eff}$.
         The result of ZEUS-EW-S 
         is shown as a cross with the error bar representing
         the total uncertainty. 
         The result in three bins with the 13 PDF parameters
         fixed to ZEUS-EW-S are shown as diamonds with experimental/fit 
         and PDF uncertainties (inner and outer error bars). 
         The band represents the SM prediction for the running of the
         effective $\sin^2\theta_W$ for the world average parameters
         as listed in PDG14. The results from LEP+SLC, CDF, D0, ATLAS,
         CMS and LHCb are at the scale of the mass of the $Z$ and horizontally
         displaced for better visibility. 
         The fixed-target experiments
         NuTeV and E158 and the determination 
         from atomic caesium, Qw(Cs), provide values
         at substantially lower scales.
}
\label{fig:run-sin}
\vfill
\end{figure}

The $\sin^2\theta_W(M_Z)$ value obtained with ZEUS-EW-S can be
converted to a value of the effective
weak mixing angle~\cite{PDG14}. 
The result is given in Table~\ref{tab:run-sin} and is shown in
Fig.~\ref{fig:run-sin} together with the SM 
prediction~\cite{run-sin-pred2} for the running 
of $\sin^2\theta_W^{\rm eff}$. 
The prediction was computed using the boson and fermion masses 
and the couplings as listed in PDG14.

An additional three fits were performed with the data
separated into three $Q^2$ bins 
from 200 to 1000, 1000 to 5000 and 5000 to 50\,000\,GeV$^2$, 
using all data available in each range.
The scales of the measurement were taken as a 
log-average $Q^2$ value of the given bin.
These bins were chosen such that the uncertainties are
about equal; cross sections for $Q^2<200$\,GeV$^2$ were found to
be insensitive to $\sin^2\theta_W$. 
%The scales were determined to be 23.8, 53.4 and 135.3\,GeV.
The PDF parameters 
were fixed to the values determined by the ZEUS-EW-S fit.
The resulting on-shell $\sin^2\theta_W$ values and the
corresponding~\cite{run-sin-pred2}    
values of $\sin^2\theta_W^{\rm eff}$  
are listed in Table~\ref{tab:run-sin} together with the values
for all data.
Also listed are the associated scales. 
Uncertainties are given for the fits themselves 
and due to the PDF parameters, model and parameterisation uncertainties
added in quadrature,
as determined by ZEUS-EW-S.
The corresponding effective $\sin^2\theta_W$ values are shown together
with the result from ZEUS-EW-S in Fig.~\ref{fig:run-sin}.
Also shown are measurements from LEP+SLC~\cite{ZLEPSLC}, D0~\cite{ZD0},
CDF~\cite{sinCDF}, CMS~\cite{sinCMS}, 
ATLAS~\cite{sinATLAS} and LHCb~\cite{sinLHCb}, 
all at the scale of the $Z$ mass, as well as a fixed-target neutrino--nucleon
measurement from
NuTeV~\cite{NuTeV}, a fixed target electron--electron measurement from
E158~\cite{sinE158}
and the result from atomic caesium~\cite{sin1Cs,sin2Cs,sin3Cs}
at lower scales. 
%at a scale of 2.4\,MeV.
This is the first time that data 
from a single experimental configuration were 
used to determine $\sin^2\theta_W$ at different
scales.
The result is compatible with the predicted running
of the effective $\sin^2\theta_W$.

The mass of the $W$ boson was determined by
a fit called ZEUS-EW-W with 13 free PDF parameters 
and, in addition, $M_W$ as a free parameter. 
The CC cross sections depend directly on $M_W$ as shown
in Eqs.~\ref{eqn:xsec_CC1} and~\ref{eqn:xsec_CC2}.
However, with $G_F$ rewritten as in Eq.~\ref{eqn:GF}, the NC data
also contribute to the fit.
The value extracted for $M_W$ is
\begin{linenomath*}\begin{equation}
\nonumber 
  M_W = 80.68~ \pm 0.28 \,{\rm {\scriptstyle{(experimental/fit)}}}~ 
          ^{+0.12}_{-0.01} \,{\rm {\scriptstyle{(model)}}} 
        ~ ^{+0.23}_{-0.01} \,{\rm {\scriptstyle{(parameterisation)}}}\,{\rm GeV}~.  
\end{equation}\end{linenomath*}
This $t$-channel determination
is in agreement with the PDG14 value of $80.385\pm0.015$\,GeV,
which is dominated by $s$-channel processes.
The result presented here is a substantial improvement compared
to a result published by H1 using HERA\,I data~\cite{ZH1}.

\begin{figure}[p]
\vfill
\begin{center}
\includegraphics[width=6in]{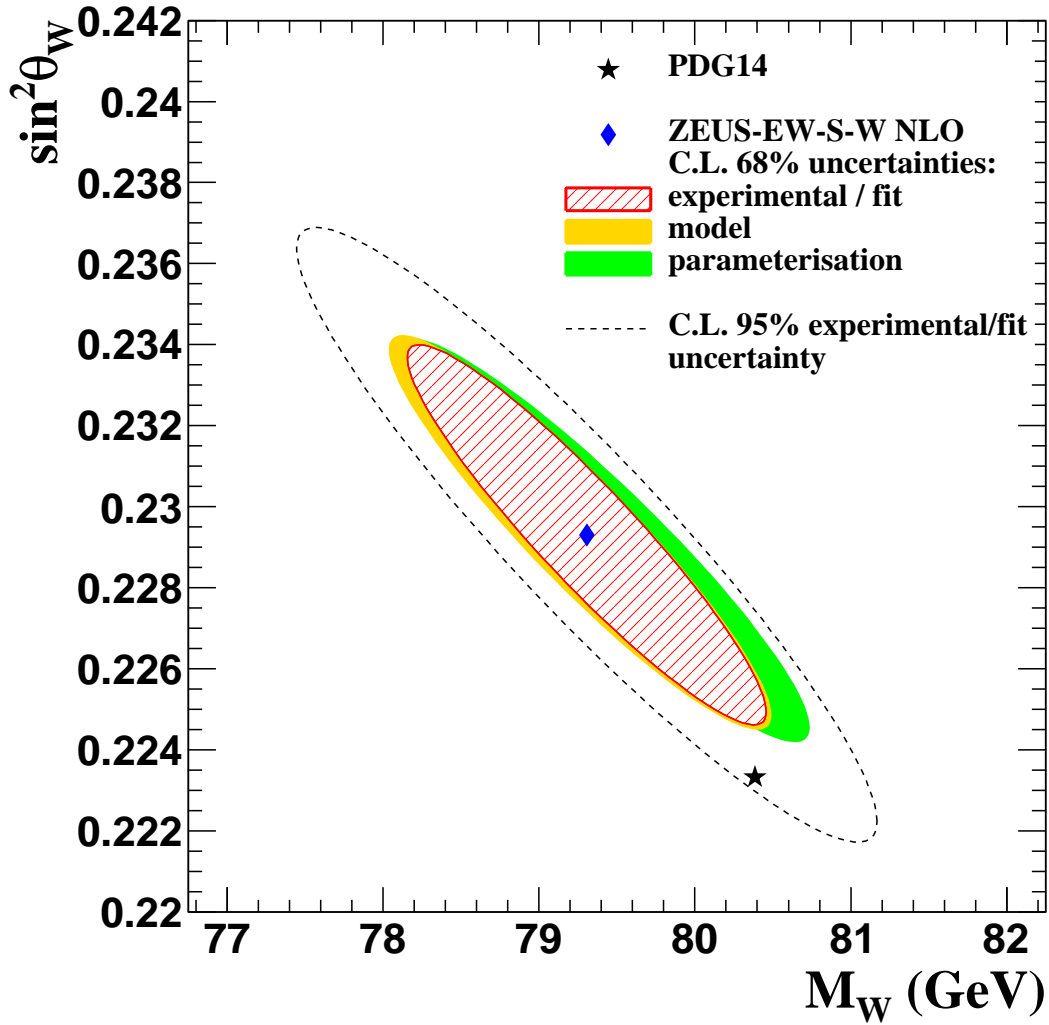}
\end{center}
\caption{The cumulative  
         68\,\%\,C.L. contour for ($M_W,\sin^2\theta_W$) 
         for the ZEUS-EW-S-W fit with experimental/fit,
         model and parameterisation uncertainties plotted 
         separately and the 95\,\%\,C.L. contour with
         experimental/fit uncertainties.
         Also shown is the world average from PDG14.
}
\label{fig:MW-sin}
\vfill
\end{figure}

Finally, a fit ZEUS-EW-S-W was performed with 13 free PDF parameters and 
both $\sin^2\theta_W$ and $M_W$  
as free parameters and with $G_F$ rewritten as described in Eq.~\ref{eqn:GF}.
The resulting values are
\begin{eqnarray}
\nonumber
  \sin^2\theta_W &=& 0.2293~ \pm 0.0031 \,{\rm {\scriptstyle{(experimental/fit)}}}~ 
                             ^{+0.0005}_{-0.0001} \,{\rm {\scriptstyle{(model)}}} 
                           ~ ^{+0.0003}_{-0.0001} \,{\rm {\scriptstyle{(parameterisation)}}}~, \\ 
\nonumber
   ~~  \\
\nonumber
  M_W &=& 79.30~ \pm 0.76 \,{\rm {\scriptstyle{(experimental/fit)}}}~ 
            ^{+0.38}_{-0.08} \,{\rm {\scriptstyle{(model)}}} 
          ~ ^{+0.48}_{-0.10} \,{\rm {\scriptstyle{(parameterisation)}}}\,{\rm GeV}~.  
\end{eqnarray}
The uncertainties on $\sin^2\theta_W$ and $M_W$ are 
substantially larger than for the
determination through ZEUS-EW-S and ZEUS-EW-W.
The values are compatible within these uncertainties.
The correlation between the EW parameters and the PDF parameters is small.
The correlation between $M_W$ and $\sin^2\theta_W$ is $-0.930$.
The 68\,\%\,C.L. contour in the $(M_W$, $\sin^2\theta_W)$ plane 
with experimental/fit, model and parameterisation
uncertainties plotted separately
is shown in Fig.~\ref{fig:MW-sin}.
Also shown is the 95\,\%\,C.L. contour with experimental/fit
uncertainties only.
The world average from PDG14 is shown as a reference.
The values for ($\sin^2\theta_W$,$M_W$) are within 2 sigma of 
the world average.

\section{Summary and conclusions}

A combined QCD and electroweak analysis was performed
based on all HERA $ep$ inclusive scattering data, exploiting the
beam polarisation for ZEUS data taken during the
years 2004 to 2007 during the HERA\,II period with a centre-of-mass
energy of 318\,GeV. 
The kinematic range of these ZEUS HERA\,II data
is $185 < Q^2 < 51\,200$\,GeV$^2$, $0.0063 < x_{\rm Bj} < 0.75$ for NC
and $200 < Q^2 < 60\,000$\,GeV$^2$, $0.0078  < x_{\rm Bj} < 1.0$ for CC
interactions.

The couplings of the $Z$ boson to $u$- and $d$-type quarks were
determined by a QCD plus EW fit with 13 parameters for the PDFs
and 4 parameters for the $Z$ couplings.
The resulting set of PDFs 
is compatible with a 13-parameter QCD-only fit and
HERAPDF2.0. 
The correlations between the PDF and coupling parameters are small.

The results for the axial-vector and vector coupling of the $Z$ boson 
to  $u$- and $d$-type quarks are
\begin{eqnarray}
\nonumber
a_u & = & +0.50~ ^{+0.09}_{-0.05} \,{\rm {\scriptstyle{(experimental/fit)}}}~ 
        ^{+0.04}_{-0.02}\,{\rm {\scriptstyle{(model)}}}~ 
        ^{+0.08}_{-0.01}\,{\rm {\scriptstyle{(parameterisation)}}} ~,\\
\nonumber
a_d & = & -0.56~ ^{+0.34}_{-0.14}\,{\rm {\scriptstyle{(experimental/fit)}}}~ 
        ^{+0.11}_{-0.05}\,{\rm {\scriptstyle{(model)}}}~ 
        ^{+0.20}_{-0.00}\,{\rm {\scriptstyle{(parameterisation)}}} ~,\\
\nonumber
v_u & = & +0.14~ ^{+0.08}_{-0.08}\,{\rm {\scriptstyle{(experimental/fit)}}}~ 
        ^{+0.01}_{-0.02}\,{\rm {\scriptstyle{(model)}}}~
        ^{+0.00}_{-0.03}\,{\rm {\scriptstyle{(parameterisation)}}}  ~,\\
\nonumber
v_d & = & -0.41~ ^{+0.24}_{-0.16}\,{\rm {\scriptstyle{(experimental/fit)}}}~ 
        ^{+0.04}_{-0.07}\,{\rm {\scriptstyle{(model)}}}~ 
        ^{+0.00}_{-0.08}\,{\rm {\scriptstyle{(parameterisation)}}} ~.
\end{eqnarray}

The values of $M_W$ and $\sin^2\theta_W$ in the on-shell scheme were extracted
with (13+1)-parameter fits.
The value extracted for $M_W$ is
\begin{linenomath*}\begin{equation}
\nonumber 
  M_W = 80.68~ \pm 0.28 \,{\rm {\scriptstyle{(experimental/fit)}}}~ 
           ^{+0.12}_{-0.01}\,{\rm {\scriptstyle{(model)}}} 
         ~ ^{+0.23}_{-0.01}\,{\rm {\scriptstyle{(parameterisation)}}}\,{\rm GeV}~.  
\end{equation}\end{linenomath*}

The on-shell value of $\sin^2\theta_W$ was determined as
\begin{linenomath*}\begin{equation}
\nonumber 
  \sin^2\theta_W = 0.2252~ \pm 0.0011 \,{\rm {\scriptstyle{(experimental/fit)}}}~ 
                 ^{+0.0003}_{-0.0001} \,{\rm {\scriptstyle{(model)}}} 
               ~ ^{+0.0007}_{-0.0001} \,{\rm {\scriptstyle{(parameterisation)}}}~.  
\end{equation}\end{linenomath*}

The determination
of  $\sin^2\theta_W$ is competitive with results obtained in the 
neutrino sector.
In addition, the data were subdivided such that values 
of the effective $\sin^2\theta_W^{\rm eff}$ 
for three different values of the scale could be determined.
The values of $\sin^2\theta_W$ and $M_W$ as well as    
of the couplings of the $Z$ boson are in agreement with
Standard Model expectations. 
The values of the axial-vector and vector couplings of the
$Z$ boson to $u$-type quarks presented in this paper are the most precise
determination published by a single collaboration.

\section{Acknowledgements}
\Zacknowledge

\vfill\eject

%------------------------------------------------------------------------------
%       Bibliography
%------------------------------------------------------------------------------
{
\def\bibname{\Large\bf References}
\def\refname{\Large\bf References}
\pagestyle{plain}
\ifzeusbst
  \bibliographystyle{l4z_default}
\fi
\ifzdrftbst
  \bibliographystyle{l4z_draft}
\fi
\ifzbstepj
  \bibliographystyle{l4z_epj}
\fi
\ifzbstnp
  \bibliographystyle{l4z_np}
\fi
\ifzbstpl
  \bibliographystyle{l4z_pl}
\fi
{\raggedright
\bibliography{DESY-16-039.bib}}

\providecommand{\etal}{et al.\xspace}
\providecommand{\coll}{Collaboration}
\catcode`\@=11
\def\@bibitem#1{%
\ifmc@bstsupport
  \mc@iftail{#1}%
    {;\newline\ignorespaces}%
    {\ifmc@first\else.\fi\orig@bibitem{#1}}
  \mc@firstfalse
\else
  \mc@iftail{#1}%
    {\ignorespaces}%
    {\orig@bibitem{#1}}%
\fi}%
\catcode`\@=12
\begin{mcbibliography}{10}

\bibitem{MANDY}
A.M. Cooper-Sarkar and R. Devenish,
\newblock {\em Deep Inelastic Scattering}.
\newblock Oxford University Press, {(2011)}\relax
\relax
\bibitem{rmp:71:1275}
H.~Abramowicz and A.~Caldwell,
\newblock Rev.\ Mod.\ Phys.{} 71~(1999)~1275\relax
\relax
\bibitem{HERAPDF20}
ZEUS and H1 Collaborations, H. Abramovicz \etal,
\newblock Eur.\ Phys.\ J.{} C~75~(2015)~580\relax
\relax
\bibitem{ZEUS2NCe}
ZEUS \coll, S. Chekanov \etal,
\newblock Eur. Phys. J.{} C 62~(2009)~625\relax
\relax
\bibitem{ZEUS2NCp}
ZEUS \coll, H. Abramowicz \etal,
\newblock Phys. Rev.{} D 87~(2013)~052014\relax
\relax
\bibitem{ZEUS2CCe}
ZEUS \coll, S. Chekanov \etal,
\newblock Eur. Phys. J.{} C 61~(2009)~223\relax
\relax
\bibitem{ZEUS2CCp}
ZEUS \coll, H. Abramowicz \etal,
\newblock Eur. Phys. J.{} C 70~(2010)~945\relax
\relax
\bibitem{H1allhQ2}
H1 Collaboration, F.D.~Aaron \etal,
\newblock JHEP{} 1209~(2012)~061\relax
\relax
\bibitem{H1FL1}
H1 Collaboration, V.~Andreev \etal,
\newblock Eur.\ Phys.\ J.{} C 74~(2014)~2814\relax
\relax
\bibitem{H1FL2}
H1 Collaboration, F.D.~Aaron \etal,
\newblock Eur.\ Phys.\ J.{} C 71~(2011)~1579\relax
\relax
\bibitem{Adloff:1999ah}
H1 Collaboration, C.~Adloff \etal,
\newblock Eur.\ Phys.\ J.{} C 13~(2000)~609\relax
\relax
\bibitem{Adloff:2000qj}
H1 Collaboration, C.~Adloff \etal,
\newblock Eur.\ Phys.\ J.{} C 19~(2001)~269\relax
\relax
\bibitem{Adloff:2003uh}
H1 Collaboration, C.~Adloff \etal,
\newblock Eur.\ Phys.\ J.{} C 30~(2003)~1\relax
\relax
\bibitem{Collaboration:2009kv}
H1 Collaboration, F.D.~Aaron \etal,
\newblock Eur.\ Phys.\ J.{} C 64~(2009)~561\relax
\relax
\bibitem{Collaboration:2009bp}
H1 Collaboration, F.D.~Aaron \etal,
\newblock Eur.\ Phys.\ J.{} C 63~(2009)~625\relax
\relax
\bibitem{ZEUSFL}
ZEUS \coll, H. Abramowicz \etal,
\newblock Phys. Rev.{} D 90~(2014)~072002\relax
\relax
\bibitem{Breitweg:1998dz}
ZEUS \coll, J.~Breitweg \etal,
\newblock Eur. Phys. J.{} C 7~(1999)~609\relax
\relax
\bibitem{zeuscc97}
ZEUS \coll, J.~Breitweg \etal,
\newblock Eur. Phys. J.{} C 12~(2000)~411.
\newblock \protect{[Erratum-ibid. C {\bf 27}, 305 (2003)}\relax
\relax
\bibitem{Chekanov:2001qu}
ZEUS \coll, S. Chekanov \etal,
\newblock Eur. Phys. J.{} C 21~(2001)~443\relax
\relax
\bibitem{Chekanov:2002zs}
ZEUS \coll, S. Chekanov \etal,
\newblock Phys. Lett.{} B 539~(2002)~197.
\newblock \protect{[Erratum-ibid. B {\bf 552}, 308 (2003)]}\relax
\relax
\bibitem{Chekanov:2002ej}
ZEUS \coll, S. Chekanov \etal,
\newblock Eur. Phys. J.{} C 28~(2003)~175\relax
\relax
\bibitem{Chekanov:2003vw}
ZEUS \coll, S. Chekanov \etal,
\newblock Eur.\ Phys.\ J.{} C 32~(2003)~1\relax
\relax
\bibitem{Chekanov:2003yv}
ZEUS \coll, S. Chekanov \etal,
\newblock Phys. Rev.{} D 70~(2004)~052001\relax
\relax
\bibitem{PDG14}
{ K.A.~Olive \etal (Particle Data Group)},
\newblock Chinese Physics{} C 38~(2014)~090001\relax
\relax
\bibitem{sovpdo:8:1203}
A.A.~Sokolov~and~I.M.~Ternov,
\newblock Sov.\ Phys.\ Dokl.{} 8~(1964)~1203\relax
\relax
\bibitem{Baier:1969hw}
V.N. Baier and V.A. Khoze,
\newblock Sov. J. Nucl. Phys.{} 9~(1969)~238\relax
\relax
\bibitem{nim:a329:79}
D.P.~Barber \etal,
\newblock Nucl.\ Inst.\ Meth.{} A~329~(1993)~79\relax
\relax
\bibitem{nim:a479:334}
M.~Beckmann \etal,
\newblock Nucl.\ Inst.\ Meth.{} A~479~(2002)~334\relax
\relax
\bibitem{Polanew}
T. Behnke \etal,
\newblock Preprint \mbox{arXiv:1201.2894}, 2012.
\newblock DESY-11-259\relax
\relax
\bibitem{Zlumi3}
L.~Adamczyk \etal,
\newblock Nucl. Inst. Meth.{} A 744~(2014)~80\relax
\relax
\bibitem{Gribov:1972ri}
V.N.~Gribov and L.N.~Lipatov,
\newblock Sov.\ J.\ Nucl.\ Phys.{} 15~(1972)~438\relax
\relax
\bibitem{Gribov:1972rt}
V.N.~Gribov and L.N.~Lipatov,
\newblock Sov.\ J.\ Nucl.\ Phys.{} 15~(1972)~675\relax
\relax
\bibitem{Lipatov:1974qm}
L.N.~Lipatov,
\newblock Sov.\ J.\ Nucl.\ Phys.{} 20~(1975)~94\relax
\relax
\bibitem{Dokshitzer:1977sg}
Y.L.~Dokshitzer,
\newblock Sov.\ Phys.\ JETP{} 46~(1977)~641\relax
\relax
\bibitem{Altarelli:1977zs}
G.~Altarelli, G.~Parisi,
\newblock Nucl.\ Phys.\ B{} 126~(1977)~298\relax
\relax
\bibitem{HERAIcombi}
ZEUS and H1 Collaborations, F.D.~Aaron \etal,
\newblock JHEP{} 1001~(2010)~109,
\newblock {and references therein}\relax
\relax
\bibitem{HERAFitter}
S. Alekhin \etal,
\newblock Eur.\ Phys.\ J.{} C 75~(2015)~304\relax
\relax
\bibitem{Martin:2009iq}
A.D. Martin \etal,
\newblock Eur. Phys. J.{} C 63~(2009)~189\relax
\relax
\bibitem{cpc:69:155}
A.~Kwiatkowski, H.~Spiesberger and H.-J.~M\"ohring,
\newblock Comp.\ Phys.\ Comm.{} 69~(1992)~155.
\newblock Also in {\it Proc.\ Workshop Physics at HERA}, eds. W.~Buchm\"{u}ller
  and G.Ingelman, (DESY, Hamburg, 1991)\relax
\relax
\bibitem{proc:hera:1991:1419}
G.A.~Schuler and H.~Spiesberger,
\newblock {\em Proc.\ Workshop on Physics at {HERA}}, W.~Buchm\"uller and
  G.~Ingelman~(eds.), Vol.~3, p.~1419.
\newblock Hamburg, Germany, DESY (1991)\relax
\relax
\bibitem{Spiesberger:95}
H.~Spiesberger,
\newblock {\em Proc.\ Workshop on Future Physics at {HERA}}, G.~Ingelman,
  A.~De~Roeck and R.~Klanner~(eds.).
\newblock Hamburg, Germany, DESY (1995)\relax
\relax
\bibitem{Bardin:89}
D.Yu.~Bardin \etal,
\newblock Z.\ Phys.\ C{} 42~(1989)~679\relax
\relax
\bibitem{Hector}
A.~Arbuzov \etal,
\newblock Comput.\ Phys.\ Commun.{} 94~(1996)~128\relax
\relax
\bibitem{ZLEPSLC}
ALEPH, DELPHI, L3 and OPAL Collaborations, SLD \coll, (LEP Electroweak Working
  Group, SLD Electroweak and Heavy Flavor Groups), G.~Abbiendi \etal,
\newblock Phys. Rept.{} 427~(2006)~257\relax
\relax
\bibitem{ZD0Old}
D0 \coll, V. Abazov \etal,
\newblock Phys. Rev.{} D 84~(2011)~012007\relax
\relax
\bibitem{ZCDF}
CDF \coll, D. Acosta \etal,
\newblock Phys. Rev.{} D 71~(2005)~052002\relax
\relax
\bibitem{ZH1}
H1 \coll, A. Aktas \etal,
\newblock Phys. Lett.{} B 632~(2006)~35\relax
\relax
\bibitem{NuTeV}
NuTeV \coll, G.P.~Zeller \etal,
\newblock Phys.\ Rev.\ Lett.{} 88~(2002)~091802.
\newblock Erratum in Phys.\ Rev.~Lett.~{90} (2003) 239902\relax
\relax
\bibitem{run-sin-pred2}
A. Czarnecki and W.J. Marciano,
\newblock Int. J. Mod. Phys.~A{} 15~(2000)~2365\relax
\relax
\bibitem{ZD0}
D0 \coll, V. Abazov \etal,
\newblock Phys. Rev. Lett.{} 115~(2015)~041801\relax
\relax
\bibitem{sinCDF}
CDF \coll, T. Aaltonen \etal,
\newblock Phys. Rev.~D{} 88~(2013)~072002.
\newblock Erratum in Phys. Rev. D{} 88~(2013)~079905\relax
\relax
\bibitem{sinCMS}
CMS \coll, S. Chatrchyan \etal,
\newblock Phys. Rev.~D{} 84~(2011)~112002\relax
\relax
\bibitem{sinATLAS}
ATLAS \coll, G. Aad \etal,
\newblock JHEP{} 1009~(2015)~049\relax
\relax
\bibitem{sinLHCb}
LHCb \coll, R. Aaij \etal,
\newblock JHEP{} 1511~(2015)~190\relax
\relax
\bibitem{sinE158}
E158 \coll, P.L.~Anthony \etal,
\newblock Phys. Rev. Lett.{} 95~(2005)~081601\relax
\relax
\bibitem{sin1Cs}
S.C. Bennett and C.E. Wieman,
\newblock Phys. Rev. Lett.{} 82~(1999)~2484\relax
\relax
\bibitem{sin2Cs}
C.S. Wood \etal,
\newblock Science{} 275~(1997)~1759\relax
\relax
\bibitem{sin3Cs}
J. Guena, M. Lintz and M.A. Bouchiat,
\newblock Phys. Rev.~A{} 71~(2005)~042108\relax
\relax
\end{mcbibliography}
%{\raggedright
%\bibliography{syn.bib,%
%              l4z_articles.bib,%
%              l4z_books.bib,%
%              l4z_conferences.bib,%
%              l4z_h1.bib,%
%              l4z_misc.bib,%
%              l4z_old.bib,%
%              l4z_preprints.bib,%
%              l4z_replaced.bib,%
%              l4z_temporary.bib,%
%              l4z_zeus.bib,%
%              desy-15-039.bib,%
%              new.bib}}
}
\vfill\eject
%------------------------------------------------------------------------------
%       Tables
%------------------------------------------------------------------------------
\newpage

\begin{table}
\begin{center}
\begin{scriptsize}
\begin{tabular}{|lr|ll|rr|c|r|r|c|c|}
\hline
\multicolumn{2}{|c|}{Data Set} &
\multicolumn{2}{c|}{$x_{\rm Bj}$} &
\multicolumn{2}{c|}{$Q^2 [$GeV$^2$]} & $e^+/e^-$ & points & 
\multicolumn{1}{ c|}{${\cal L}$} 
& $P_e$  &  Ref. \\
process & year & from & to & from & to & &
& \multicolumn{1}{ c|}{pb$^{-1}$}   &   &  \\
\hline
NC  & $06$-$07$ & $0.0063$   &$0.75$   &$185$   &$50000$   & $e^+p$ & 90
& 78.8$\pm$1.4  &  $+$0.316 $\pm$ 0.013    & \cite{ZEUS2NCp}\\
&  &  &  &   &   &  & 90
& 56.7$\pm$1.1 &  $-$0.353 $\pm$    0.014    & \\
\hline
CC  & $06$-$07$ & $0.0078$   &$1.00$   &$280$   &$50000$  & $e^+p$   & 35
&  ~75.8$\pm$1.4 &  $+$0.327 $\pm$ 0.012  & \cite{ZEUS2CCp}\\
&  &  &  &    &   &  & 35
& ~56.0$\pm$1.1  &  $-$0.358 $\pm$    0.014     & \\
\hline
NC  & $05$-$06$ & $0.0063$  &$0.75$   &$185$   &$51200$   & $e^-p$ & 90
&  ~71.2$\pm$1.3 & $+$0.289 $\pm$ 0.011  &  \cite{ZEUS2NCe}\\
&  &  &  &    &   &  & 90
& ~98.7$\pm$1.8  &  $-$0.262 $\pm$    0.011     & \\
\hline
CC  & $04$-$06$ & $0.010$   &$1.00$   &$200$   &$60000$   & $e^-p$ & 34
& ~71.0$\pm$1.3   &  $+$0.296 $\pm$ 0.011  & \cite{ZEUS2CCe}\\
&  &  &  &   &   &  & 37
& 104.0$\pm$1.9   &  $-$0.267 $\pm$ 0.011  & \\
\hline 
\end{tabular}
\end{scriptsize}
\end{center}
\caption{\label{tab:pol}The four ZEUS data sets, for which polarisation was
taken into account. 
%This doubles the number of cross sections for these
%data sets.}
}
\end{table}
%Lumi: 1.8% fuer alle ausser e+p letzte eproide P=-.36

\begin{sidewaystable}
\begin{center}
\begin{scriptsize}\renewcommand\arraystretch{1.1}
\begin{tabular}[H]{ | c | r r r r r r r r r r r r r r r r r |}
\hline
\hline
Parameters & {\it xg: B}       & {\it xg: C}   & {\it xg: A$^{'}$}    & {\it xg: B$^{'}$}    & {\it xu$_{v}$: B}   & {\it xu$_{v}$: C}  & {\it xu$_{v}$: E}    & {\it xd$_{v}$: B}    & {\it xd$_{v}$: C}  & {\it x$\bar{U}$: C} & {\it x$\bar{D}$: A}  & {\it x$\bar{D}$: B} & {\it x$\bar{D}$: C} & $a_{u}$ & $a_{d}$ & $v_{u}$ &  $v_{d}$ \rule{0pt}{3ex}  \\
\hline \rule{0pt}{3ex}
{\it xg: B}         & 1.000  & $-$0.014  & $-$0.449  & 0.824  & $-$0.216  & 0.172  & 0.250  & $-$0.084  & $-$0.085  & $-$0.098  & $-$0.107  & $-$0.136   & 0.046   & 0.025   & 0.003   & 0.015   & 0.018 \rule{0pt}{3ex} \\
{\it xg: C}         & $-$0.014  & 1.000  & 0.831  & 0.457  & 0.341  & $-$0.373  & $-$0.550  & 0.010  & 0.296  & $-$0.018  & $-$0.082  & $-$0.103   & $-$0.434   & 0.105   & 0.095   & $-$0.098   & $-$0.111\\
{\it xg: A$^{'}$}      & $-$0.449  & 0.831  & 1.000  & 0.120  & 0.548  & $-$0.404  & $-$0.629  & 0.233  & 0.274  & 0.159  & 0.081  & 0.072   & $-$0.148   & $-$0.052   & 0.000   & $-$0.043   & $-$0.054\\
{\it xg: B$^{'}$}      & 0.824  & 0.457  & 0.120  & 1.000  & 0.106  & $-$0.037  & $-$0.082  & 0.075  & 0.047  & 0.043  & 0.011  & $-$0.014   & 0.012   & $-$0.029   & $-$0.011   & $-$0.001   & $-$0.002\\
{\it xu$_{v}$: B}        & $-$0.216  & 0.341  & 0.548  & 0.106  & 1.000  & $-$0.409  & $-$0.774  & 0.465  & $-$0.086  & 0.690  & 0.476  & 0.395   & 0.439   & $-$0.360   & $-$0.178   & 0.079   & 0.070\\
{\it xu$_{v}$: C}        & 0.172  & $-$0.373  & $-$0.404  & $-$0.037  & $-$0.409  & 1.000  & 0.828  & $-$0.297  & $-$0.235  & $-$0.188  & $-$0.095  & $-$0.069   & $-$0.040   & 0.110   & 0.029   & 0.040   & 0.028\\
{\it xu$_{v}$: E}        & 0.250  & $-$0.550  & $-$0.629  & $-$0.082  & $-$0.774  & 0.828  & 1.000  & $-$0.296  & $-$0.066  & $-$0.363  & $-$0.170  & $-$0.117   & $-$0.092   & 0.192   & 0.087   & $-$0.023   & $-$0.017\\
{\it xd$_{v}$: B}        & $-$0.084  & 0.010  & 0.233  & 0.075  & 0.465  & $-$0.297  & $-$0.296  & 1.000  & 0.518  & 0.405  & 0.350  & 0.291   & 0.673   & $-$0.335   & $-$0.134   & 0.038   & 0.021\\
{\it xd$_{v}$: C}        & $-$0.085  & 0.296  & 0.274  & 0.047  & $-$0.086  & $-$0.235  & $-$0.066  & 0.518  & 1.000  & $-$0.137  & $-$0.186  & $-$0.193   & $-$0.139   & 0.110   & 0.128   & $-$0.101   & $-$0.128\\
{\it x$\bar{U}$: C}      & $-$0.098  & $-$0.018  & 0.159  & 0.043  & 0.690  & $-$0.188  & $-$0.363  & 0.405  & $-$0.137  & 1.000  & 0.673  & 0.635   & 0.329   & $-$0.320   & $-$0.137   & 0.055   & 0.052\\
{\it x$\bar{D}$: A}      & $-$0.107  & $-$0.082  & 0.081  & 0.011  & 0.476  & $-$0.095  & $-$0.170  & 0.350  & $-$0.186  & 0.673  & 1.000  & 0.959   & 0.477   & $-$0.272   & $-$0.137   & 0.056   & 0.059\\
{\it x$\bar{D}$: B}      & $-$0.136  & $-$0.103  & 0.072  & $-$0.014  & 0.395  & $-$0.069  & $-$0.117  & 0.291  & $-$0.193  & 0.635  & 0.959  & 1.000   & 0.415   & $-$0.239   & $-$0.120   & 0.047   & 0.053\\
{\it x$\bar{D}$: C}      & 0.046  & $-$0.434  & $-$0.148  & 0.012  & 0.439  & $-$0.040  & $-$0.092  & 0.673  & $-$0.139  & 0.329  & 0.477  & 0.415   & 1.000   & $-$0.449   & $-$0.271   & 0.148   & 0.153\\
$a_{u}$       & 0.025  & 0.105  & $-$0.052  & $-$0.029  & $-$0.360  & 0.110  & 0.192  & $-$0.335  & 0.110  & $-$0.320  & $-$0.272  & $-$0.239   & $-$0.449   & 1.000   & 0.861   & $-$0.555   & $-$0.729\\
$a_{d}$       & 0.003  & 0.095  & 0.000  & $-$0.011  & $-$0.178  & 0.029  & 0.087  & $-$0.134  & 0.128  & $-$0.137  & $-$0.137  & $-$0.120   & $-$0.271   & 0.861   & 1.000   & $-$0.636   & $-$0.880\\
$v_{u}$       & 0.015  & $-$0.098  & $-$0.043  & $-$0.001  & 0.079  & 0.040  & $-$0.023  & 0.038  & $-$0.101  & 0.055  & 0.056  & 0.047   & 0.148   & $-$0.555   & $-$0.636   & 1.000   & 0.851\\
$v_{d}$       & 0.018  & $-$0.111  & $-$0.054  & $-$0.002  & 0.070  & 0.028  & $-$0.017  & 0.021  & $-$0.128  & 0.052  & 0.059  & 0.053   & 0.153   & $-$0.729   & $-$0.880   & 0.851   & 1.000 \\
\hline
\hline
\end{tabular}
\caption{\label{tab:matrix}The correlation matrix of all parameters 
of the ZEUS-EW-Z fit.}
\end{scriptsize}
\end{center}
\end{sidewaystable}

\newpage
\begin{table}
\renewcommand*{\arraystretch}{1.3}
\begin{center}
\begin{scriptsize}
\begin{tabular}{|l|ccc|ccc|ccc|ccc|}
\hline 
    & $a_u$ & exp & tot &  $a_d$ & exp &  tot & 
      $v_u$ & exp & tot &  $v_d$ & exp &  tot \\
\hline 
%after reading --> 2 digits 
  EW-Z & $+0.50$ & $^{+0.09}_{-0.05}$ &$ ^{+0.12}_{-0.05}$& 
         $-0.56$ & $^{+0.34}_{-0.14}$ &$ ^{+0.41}_{-0.15}$& 
         $+0.14$ & $^{+0.08}_{-0.08}$ &$ ^{+0.09}_{-0.09}$& 
         $-0.41$ & $^{+0.24}_{-0.16}$ &$ ^{+0.25}_{-0.20}$ \\
\hline
 13p & $+0.49$ & $^{+0.07}_{-0.04} $ & &
         $-0.57$ & $^{+0.30}_{-0.13} $ & &
         $+0.15$ & $^{+0.08}_{-0.08}$  & &
         $-0.40$ & $^{+0.22}_{-0.17}$  &   \\
\hline
HPDF1 & $+0.47$ & $^{+0.06}_{-0.03}$ & & 
         $-0.62$ & $^{+0.23}_{-0.11}$ & & 
         $+0.16$ & $^{+0.08}_{-0.08}$ & & 
         $-0.35$ & $^{+0.22}_{-0.19}$ & \\
\hline
% MSbar value of sin2theta W in fit
HPDF2 & $+0.49$ & $^{+0.06}_{-0.03}$ & & 
         $-0.63$ & $^{+0.24}_{-0.11}$ & & 
         $+0.15$ & $^{+0.08}_{-0.08}$ & & 
         $-0.36$ & $^{+0.22}_{-0.19}$ & \\
\hline
SM     & $+0.50$ &&& $-0.50$ &&& $+0.20$ &&& $-0.35$ && \\
\hline
\end{tabular}
\end{scriptsize}
\end{center}
\caption{\label{tab:Z}The results on the axial-vector and
vector couplings of the $Z$ boson to 
$u$- and $d$-type quarks from ZEUS-EW-Z.
Given are the experimental/fit (exp) and total (tot)
uncertainties.
Also listed are results of fits with the PDFs fixed to
ZEUS-13p and HERAPDF2.0, HPDF1 and HPDF2, for which
only the couplings of the $Z$ were free parameters. 
The HPDF1 fit was performed with the on-shell value
of $\sin^2\theta_W$ used in the fit while HPDF2
was performed with the $\sin^2\theta_W$ value used
for the extraction of HERAPDF2.0.
Also listed are
the predictions of the SM for the $a$ and $v$ couplings
in the on-shell scheme.}
\end{table}

\newpage

\begin{table}
\renewcommand*{\arraystretch}{1.3}
\begin{center}
\begin{scriptsize}
\begin{tabular}{|l|r|r|r|c|c|l|c|c|}
\hline
bin & $Q^2_{\rm min}$ & $Q^2_{\rm max}$ & scale & $\sin^2\theta_W$ & exp 
    %& PDF   
    & $\sin^2\theta_W^{\rm eff}$ & exp  & PDF  \\
    & (GeV$^2$)     & (GeV$^2$)      & (GeV) & on-shell         
    %& unc.    
    & unc. & effective         & unc.    & unc. \\
\hline
1 & 200  & 1000	& 22.3  & 0.2254 &  $\pm$0.0020 %& $^{+0.0019}_{-0.0012}$ 
                        & 0.2352 &  $\pm$0.0020 & $^{+0.0020}_{-0.0012}$ \\
\hline
2 & 1000 & 5000 & 49.9  & 0.2251 &  $\pm$0.0014 %& $^{+0.0014}_{-0.0008}$ 
                        & 0.2339 &  $\pm$0.0015 & $^{+0.0014}_{-0.0008}$ \\ 
\hline
3 & 5000 &50000 & 139.8 & 0.2240 & $\pm$0.0026	%& $^{+0.0026}_{-0.0015}$ 
                        & 0.2323 &  $\pm$0.0026	& $^{+0.0025}_{-0.0015}$ \\
\hline
\multicolumn{3}{|l|}{All Data} 
                & $M_Z$ & 0.2252 & $\pm$0.0011	%& $^{+0.0008}_{-0.0002}$ 
                        & 0.2335 &  $\pm$0.0011 & $^{+0.0008}_{-0.0004}$ \\
\hline
\end{tabular}
\end{scriptsize}
\end{center}
\caption{\label{tab:run-sin}The on-shell and effective 
values of $\sin^2\theta_W$ as
determined for three bins in $Q^2$ and for all data.
Experimental/fit (exp) uncertainties are given as determined
by the one-parameter fits for each bin or ZEUS-EW-S, respectively;
model and parameterisation uncertainties as determined by ZEUS-EW-S
were added in quadrature and are denoted as PDF uncertainties.
They are identical for on-shell and effective values at the accuracy given.} 
\end{table}

\vfill\eject
%------------------------------------------------------------------------------
%       Figures
%

%---------------------------------------------------------------------------
% -- ZEUS EW ZC PDFs
\clearpage

\newpage
\newpage

% -------------------------------------------------------
% NC data

\newpage

%------------------------------------
% coupling contours a against v
\newpage
%------------------------------------
% coupling contours a against a
\newpage
%-comparisons to others
\newpage

%----------------
% coupling comparison bars

%-------------------------------------
% -- ZEUS EW W PDFs --> S new

% CC data------------------------------
\newpage

\newpage

\newpage

%contour MW sin2thetaW

\newpage
%contour MW sin2thetaW

\end{document}